\documentclass[12pt,preprint,longabstract]{aastex}

\shorttitle{Polarization Properties of Radio Galaxies at 1.4 GHz}
\shortauthors{Grant et.\ al.}

\begin{document}

\title{The DRAO Planck Deep Fields: the polarization properties of radio galaxies at 1.4 GHz}
\author{J.\ K. Grant, A.\ R. Taylor, J.\ M. Stil}
\affil{Institute for Space Imaging Science, University of Calgary, AB T2N 1N4, Canada}

\author{T.\ L. Landecker, R. Kothes, R.\ R. Ransom}
\affil{Dominion Radio Astrophysical Observatory, Herzberg Institute of Astrophysics, National Research Council Canada, Penticton, BC V2A 6J9, Canada}

\author{Douglas Scott}
\affil{Department of Physics and Astronomy, University of British Columbia, Vancouver, BC V6T 1Z1, Canada}

\begin{abstract}
We present results of deep polarization imaging at 1.4 GHz with the Dominion Radio Astrophysical Observatory as part of the DRAO Planck Deep Fields project.  This deep extragalactic field covers 15.16 deg$^2$ centered at $\alpha_{2000} = 16^{{\rm h}} 14^{{\rm m}}$ and $\delta_{2000} = 54\arcdeg 56\arcmin$, has an angular resolution of $42\arcsec \times 62\arcsec$ at the field center, and reaches a sensitivity of 55 $\mu$Jy beam$^{-1}$ in Stokes $I$ and 45 $\mu$Jy beam$^{-1}$ in Stokes $Q$ and $U$.  We detect 958 radio sources in Stokes $I$ of which 136 are detected in polarization.  We present the Euclidean-normalized polarized differential source counts down to 400 $\mu$Jy.  These counts indicate that sources have a higher degree of fractional polarization at fainter Stokes $I$ flux density levels than for brighter sources, confirming an earlier result.  We find that the majority of our polarized sources are steep-spectrum objects with a mean spectral index of $-0.77$, and there is no correlation between fractional polarization and spectral index.  We also matched deep field sources to counterparts in the Faint Images of the Radio Sky at Twenty Centimeters catalogue.  Of the polarized sources, $77\%$ show structure at the arc-second scale whereas only $38\%$ of the sources with no detectable polarization show such structure.  The median fractional polarization is for resolved sources is $6.8\%$, while it is $4.4\%$ for compact objects.  The polarized radio sources in our deep field are predominantly those sources which are resolved and show the highest degrees of fractional polarization, indicating that the lobe dominated structure may be the source of the highly polarized sources.  These resolved radio galaxies dominate the polarized source counts at $P_{\rm 0} = \sqrt{Q^2 + U^2} <3\,$mJy.
\end{abstract}


\section{Introduction}
Cosmic magnetism has been the focus of many scientific papers describing efforts to understand the role of magnetic fields in astrophysics.   Cosmic magnetic fields are observed through synchrotron radiation at radio wavelengths.  The percentage polarization ($\Pi = P/I$) and plane of polarization of the radiation provide information regarding the strength and direction of the magnetic field, while Faraday rotation provides directly information about the line-of-sight component of the magnetic field.

One way of studying cosmic magnetic fields is through statistical studies of the properties of polarized radio sources.  Radio source counts provide insight into the evolution of galaxies over a range of frequencies, luminosities, and galaxy types.  Total intensity (i.e. Stokes $I$) radio source counts have been well studied down to $\sim10\,\mu$Jy in concentrated areas \citep{Hopkins2003,Windhorst2003,Owen2008}.  \citet{Windhorst2003} used 1.4 GHz surveys and models to show that evolving giant elliptical galaxies and quasars dominate the radio source counts for flux densities at 1.4 GHz of $S_{1.4} \ge 0.5\,$mJy, while starburst and normal galaxies begin to dominate the source counts at $S_{1.4} \le 0.3\,$mJy.  \citet{Owen2008} produce radio source counts at 1.4 GHz over a region of the \textit{Spitzer} Wide-area InfraRed Extragalactic Survey \citep[SWIRE,][]{Lonsdale2003} down to $15\,\mu$Jy.  They note that there may be a natural confusion limit for total intensity near $1\,\mu$Jy given the finite median source size of $\sim1.2\arcsec$.  All-sky surveys like the NRAO VLA Sky Survey \citep[NVSS,][]{Condon1998} and Faint Images of the Radio Sky at Twenty Centimeters \citep[FIRST,][]{White1997} have provided 1.4 GHz source counts with high accuracy down to $2.5\,$mJy and $1\,$mJy, respectively.

Polarized radio source counts have only just begun to be studied.  The NVSS was used by \citet{Mesa2002} for a study of polarized source counts by correlating sources at 1.4 GHz from the NVSS with sources from the Green Bank 4.85 GHz catalogue \citep{Gregory1996}.  Sources with $S_{1.4} \ge 80\,$mJy and $\Pi \ge 1\%$ were found to show an increase in degree of polarization with decreasing flux density.  This trend is  particularly evident for steep spectrum sources where the median $\Pi$ increases from $\Pi \simeq1.1\% $ for $S > 800\,$mJy to $\Pi \simeq1.8\%$ or $100 < S < 200\,$mJy.  They suggest that the origin of this correlation could be a change in the source population with decreasing flux density.  Analyzing the polarization structure of the sources with $S > 500\,$mJy \citet{Mesa2002} noted that extended sources were less polarized than compact ones.  \citet{Tucci2004} confirmed the \citet{Mesa2002} result for steep spectrum sources but not for flat spectrum sources.  A possible explanation for this effect is an increase of the mean redshift with decreasing flux density.  \citet{Tucci2004} also noted that the Euclidean-normalized polarized source counts remained constant down to a flux density of $1\,$mJy.

\citet{Beck2004} extrapolated the Euclidean-normalized polarized source counts below $1\,$mJy by convolving the Stokes $I$ differential source count distribution \citep[see also][]{Hopkins2000} with a probability distribution function in fractional polarized intensity, the later constructed using polarized sources from the NVSS with $S \ge 80\,$mJy.  This model does not include a flux density dependence of fractional polarization, although they note that their extrapolation of the polarized source counts is a lower limit.  Actual measurements of polarized radio sources down to $500\,\mu$Jy were presented by \citet{Taylor2007} who found that the Euclidean-normalized polarized radio source counts remain constant down to $500\,\mu$Jy, with a continued increase in the fractional polarization of radio sources below 80 mJy.  The median $\Pi$ increases to 4.8\% for sources with $10 < S_{1.4} < 30$ mJy.  Deeper polarized source counts are required to determine if the anti-correlation of $\Pi$ with flux density continues to lower flux density levels and to explore the polarized source count statistics at the boundary where the dominant population may be expected to change from AGN to starburst galaxies.

We present a continuation of the \citet{Taylor2007} polarization study of the DRAO ELAIS N1 deep field to study the polarized source population in the European Large Area \textit{ISO} Survey North 1 region \citep[ELAIS N1,][]{Oliver2000}.  Aside from the observations at 15, 90, and $175\,\mu$m with \textit{ISO} \citep{Oliver2000}, the ELAIS N1 region has been observed at a multitude of other wavelengths.  \citet{Ciliegi1999} used the VLA in C configuration at $20\,$cm in total intensity and reached a 5$\sigma$ limit of $135\,\mu$Jy over $0.12\,$deg$^2$, the \textit{Spitzer} Wide-area InfraRed Extragalactic survey \citep[SWIRE,][]{Lonsdale2003} used IRAC at 3.6, 4.5, 5.6, and $8.0\,\mu$m and MIPS at 24, 70, and $160\,\mu$m, ground based optical imaging ($u', g', r', i', z'$) covering the SWIRE fields was done by the Issac Newton Telescope Wide Field Survey \citep[INT WFS,][]{McMahon2001}, \citet{Manners2003} observed ELAIS N1 with \textit{Chandra}, and the Giant Metrewave Radio Telescope (GMRT) was used at 610 MHz \citep{Garn2008} and 325 MHz \citep{Sirothia2009}.  The ELAIS N1 field is also covered by the Westerbork Northern Sky Survey \citep[WENSS,][]{wenss}, NVSS, and FIRST.  Combining these data produces a large database for extragalactic source studies.  By increasing both the sensitivity and coverage area of the \citet{Taylor2007} study, we are able to observe a larger number of polarized radio sources.  The total area of the observations has been increased by a factor of two, allowing us to increase the number of sources i the same flux density range as \citet{Taylor2007}.  Increasing the sensitivity of the observations allows us to investigate the polarized radio sources to the faintest flux densities to date and probe the nature of the faint polarized source population.  In section 2, we describe the Dominion Radio Astrophysical Observatory synthesis telescope (DRAO ST) observations and preparation of the polarization images.  In section 3, the method of detecting and cataloguing the DRAO Planck Deep Fields radio sources in the ELAIS N1 field is described along with the matching to counterparts in WENSS, NVSS, and FIRST.  In section 4, we describe the nature of the polarized sources in our sample.

\section{Observations and Data Processing}
\subsection{Synthesis Observations}
The Dominion Radio Astrophysical Observatory synthesis telescope \citep[DRAO ST,][]{Landecker} was used to produce a deep polarization mosaic at 1.4 GHz centered on the ELAIS N1 field, expanding upon  pilot observations reported by \citet{Taylor2007} in both solid angle and sensitivity.    The DRAO ST simultaneously observed the HI 21 cm line and 408 MHz continuum along with 1.4 GHz continuum.  The seven dish east-west interferometer observed in full polarimetry continuum mode using the four available bands of 1406.65, 1414.15, 1426.65, and 1434.15 MHz, each with a bandwidth of 7.5 MHz.  A full synthesis or each individual field requires 144 hours to complete by moving three antennas in increments of 4.3 m to provide complete sampling of the UV plane.  The shortest baseline is 12.86 m while the longest baseline is 617.18 m.  The primary beam has a FWHM of 107.2$\arcmin$ at 1.4 GHz and places the first grating lobe outside of the primary beam.  The synthesized beam has the first sidelobe at the 3\% level, while the rest of the sidelobes are less than 0.5\% of the peak of the main lobe.  The angular resolution of the DRAO ST at 1.4 GHz is $\sim60\arcsec$ and the effective beam can be fit by a two-dimensional gaussian.   However, the angular resolution of a mosaic is $b_\alpha \times b_\delta = 49\arcsec \times 49\arcsec {\rm cosec}\delta$, which varies over the mosaic.  A more detailed description of the DRAO ST can be found in \citet{Landecker}.

The observations of the DRAO ELAIS N1 deep field began in August 2004 and the final deep field observation was completed in July 2008.  Since this project began, many improvements to the interferometer have been completed.  These improvements include the installation of new low noise amplifiers and shielding fences to block ground radiation.  In 2000, the Stokes $I$ noise at the center of the individual fields was $280\,\mu$Jy beam$^{-1}$ \citep{Landecker}.  For the \citet{Taylor2007} study, the noise was improved to $210\,\mu$Jy beam$^{-1}$.  For the additional fields in our study, the rms noise level is $189\,\mu$Jy beam$^{-1}$.  

Individual fields are mosaiced together to achieve higher sensitivity and a wider field of view.  Data from each individual field are included out to a radius of $93\arcmin$ for Stokes $I$ and $75\arcmin$ for Stokes $Q$ and $U$.  The smaller radius cutoff for polarization results from the fact that the instrumental leakages outside $75\arcmin$ are expected to be large and require long integration times to measure accurately \citep{Reid2008}. The data from overlapping fields are averaged together with weights \citep{CGPS}
\begin{equation}
w_m = \left[\frac{A(\rho_m)}{\sigma_m}\right]^2\,,
\end{equation}
where $A(\rho_m)$ is the primary beam attenuation at pixel distance $\rho_m$ from the field center, and $\sigma_m$ is the individual field rms noise level.  The resulting rms noise level in the mosaic is then a function of location in the mosaic, and is given by \citep{CGPS}
\begin{equation}\label{weights}
\sigma = \left[\sum^N_{m=1}w_m\right]^{-1/2}\,.
\end{equation}

\begin{figure}
\epsscale{1.0}
\plotone{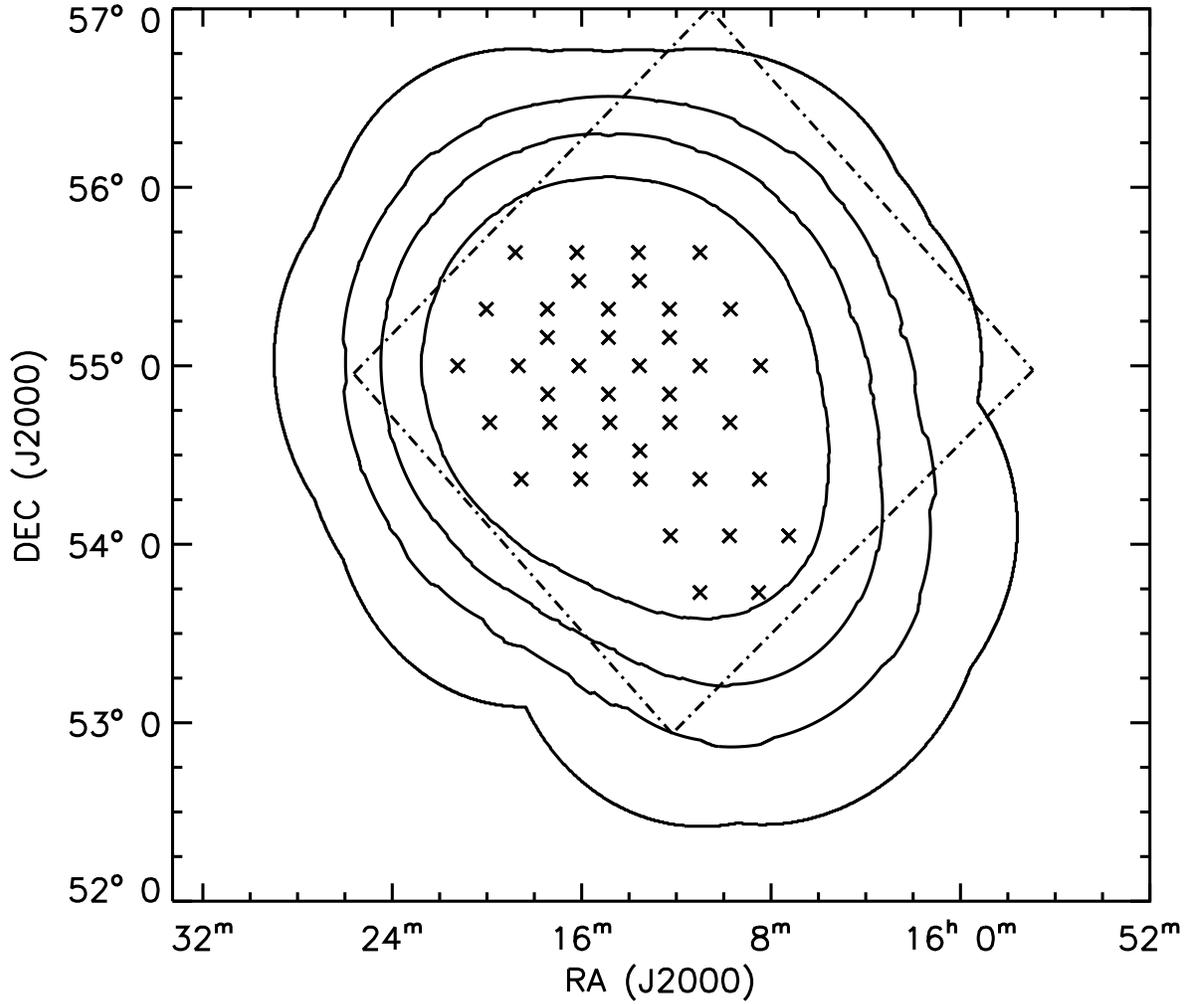}
\caption{The DRAO ELAIS N1 deep field polarization survey area (---) with respect to the SWIRE survey (-- $\cdot$ --).  The contours indicate the sensitivity of the DRAO ELAIS N1 survey at 2, 3, 5 and 20 times the theoretical noise level.  The $\times$ symbols indicate the location of the 40 field pointings.}\label{fieldcen}
\end{figure}

The DRAO ELAIS N1 deep field mosaic is comprised of 40 individual fields.  The first 30 fields were placed in a triangular pattern and separated by $22\arcmin$ to provide an rms noise level of $56\,\mu$Jy beam$^{-1}$ per polarization.  The additional 10 fields were placed in the inner region of the 30 fields in order to achieve a sensitivity of $45\,\mu$Jy beam$^{-1}$ in Stokes $Q$ and $U$. Figure \ref{fieldcen} shows the DRAO ELAIS N1 deep field pointings, for comparison the area covered by the SWIRE survey is also shown.

The DRAO ST is equipped to receive both left- and right-handed circularly polarized radiation.  The polarimetric imaging methods described in the Canadian Galactic Plane Survey \citep[CGPS,][]{CGPS, Landecker2009}, with $uv$-plane based polarization leakage correction described by \citet{Reid2008} and \citet{Ransom2008}, were used to create the deep polarization mosaic of the ELAIS N1 field.  The instrumental polarization correction is less than 0.5\% in the central area of the mosaic and can be as large as 1\% at the edges of the mosaic \citep{CGPS}.  The polarized sources presented in this paper are contained in the area of the mosaic where the instrumental polarization is less than 0.5\%.  The polarization calibrators used for the observations were 3C147 and 3C295 which are both unpolarized and unresolved.  These two calibrators were observed between each 12 hour observation.  Additionally, the polarization angle calibrator, 3C286, was observed once every four days.  

\begin{figure}
\epsscale{1.0}
\includegraphics[width=0.5\textwidth]{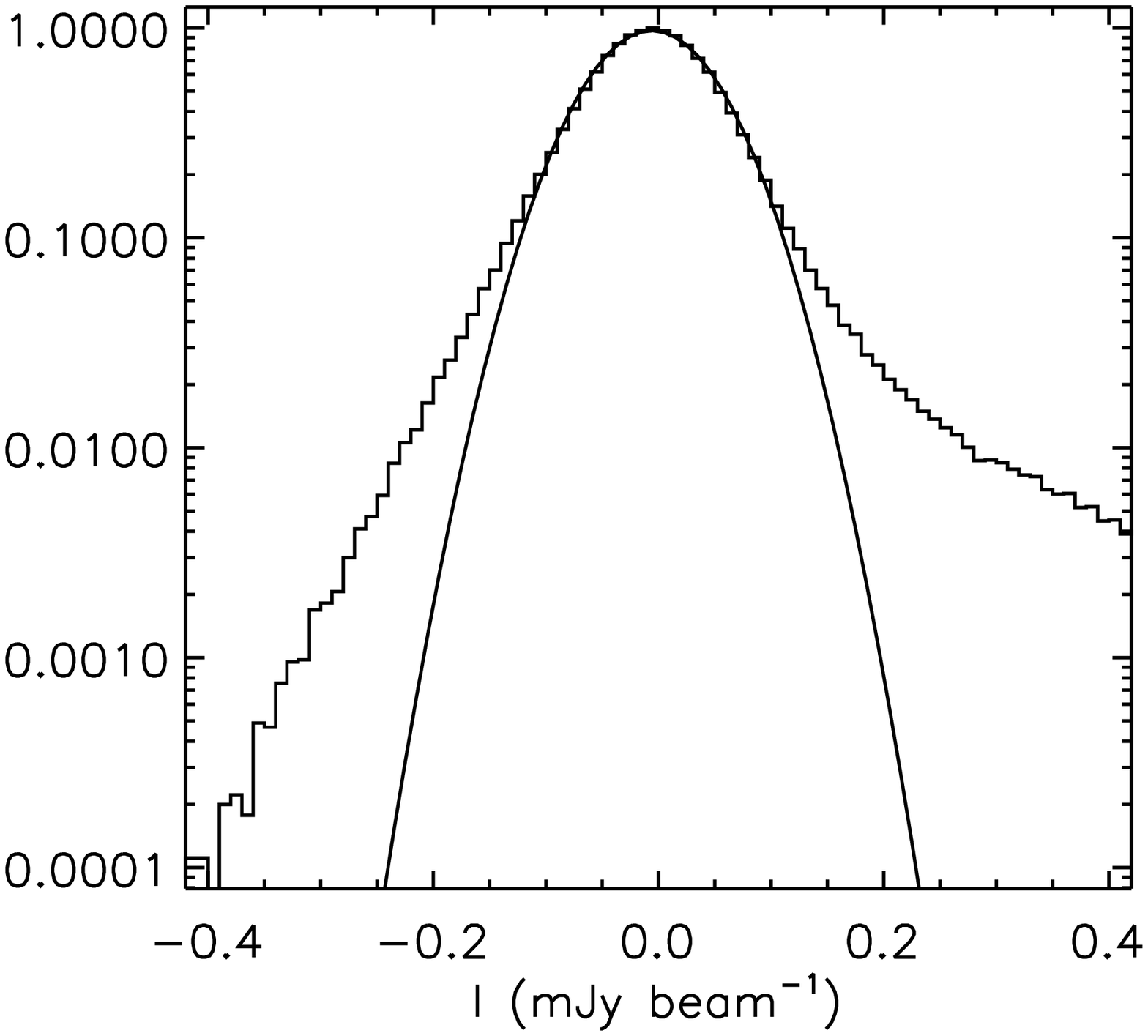}
\includegraphics[width=0.5\textwidth]{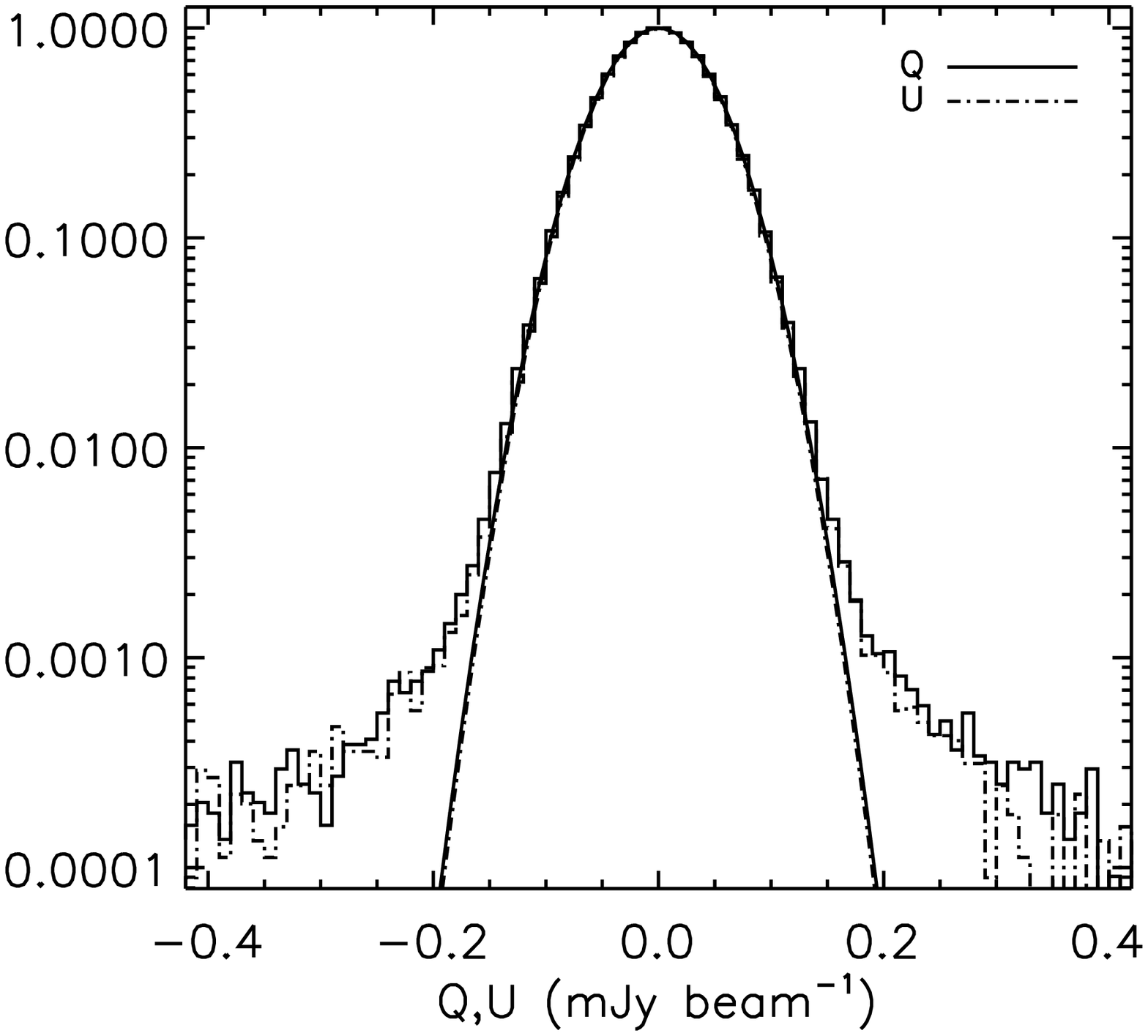}
\caption{The distribution of flux densities in Stokes $I$ (left) and $Q$ and $U$ (right) of the uniform noise mosaics.  Gaussian fits to the distribution were used to measure the noise value of $55\,\mu$Jy beam$^{-1}$ for Stokes $I$ and $45\,\mu$Jy beam$^{-1}$ for Stokes $Q$ and $U$.}\label{noisegauss}
\end{figure}

The Sun also produces polarized interference, as the antenna sidelobes are highly polarized.  The solar interference was removed by creating images of the Sun during the observations and removing the response from the visibility data.  Terrestrial radio frequency interference is also polarized since these signals arise from stationary sources that appear as a source at the North Celestial Pole, because the relative position of these interfering sources relative to the telescopes remains always constant.  By imaging the NCP, much of the interference can be removed from the visibility data.  For some subsets of the data, there was more complex interference in which case the visibility data were flagged.  The CLEANing process \citep{Clark1980} and modcal, a direction-dependent self-calibration \citep{Willis1999}, were implemented to remove the majority of the artifacts from the mosaics.  There are a few artifacts left and the sensitivity is limited by thermal noise levels, see Figure \ref{noisegauss}.

\subsection{The Mosaics}
The DRAO ELAIS N1 deep field mosaics were produced in Stokes $I$, $Q$ and $U$, centered on $\alpha_{2000} = 16^{{\rm h}} 14^{{\rm m}}$ and $\delta_{2000} = 54\arcdeg 56\arcmin$, and have an angular resolution which varies over the mosaic according to $b_\alpha \times b_\delta = 49\arcsec \times 49\arcsec {\rm cosec}\delta$.  The center of the mosaic has a measured angular resolution of 49$\arcsec \times 62\arcsec$. The total intensity image covers 19.68 deg$^2$ while the polarization mosaics covers 15.16 deg$^2$ as there is a difference in the primary beam cutoff (see section 2.1).  The mosaicing process produces the lowest noise in the central region of the mosaic where the sum of the weights are the largest.  By removing the weight distribution from the mosaic according to equation [\ref{weights}], the mosaic is made to have a rms noise value for each pixel equal to the rms noise level for the maximum sum of weights.  This uniform noise mosaic is used for subsequent analysis.

\begin{figure}
\epsscale{1.0}
\includegraphics[width=0.5\textwidth]{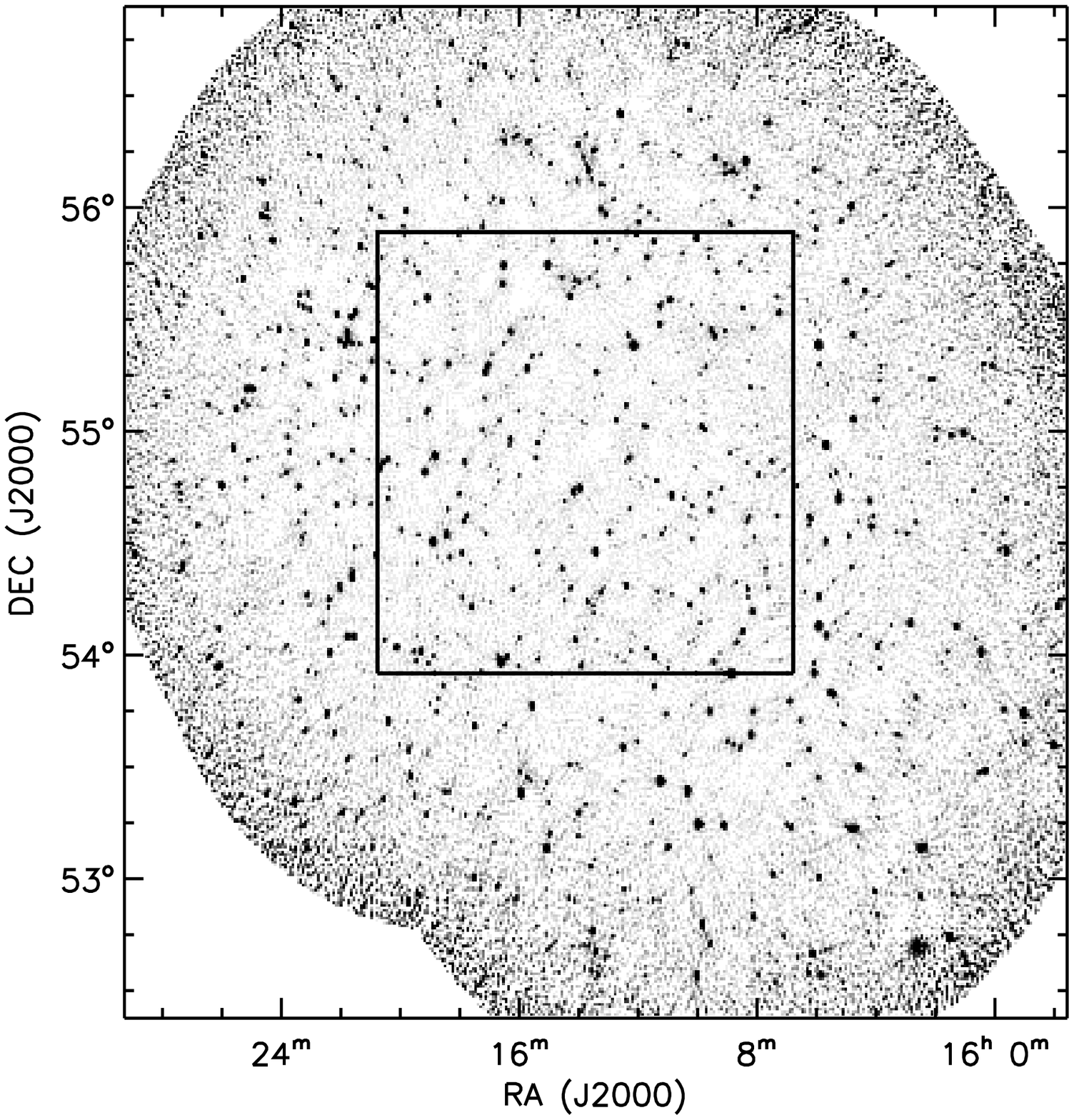}
\includegraphics[width=0.5\textwidth]{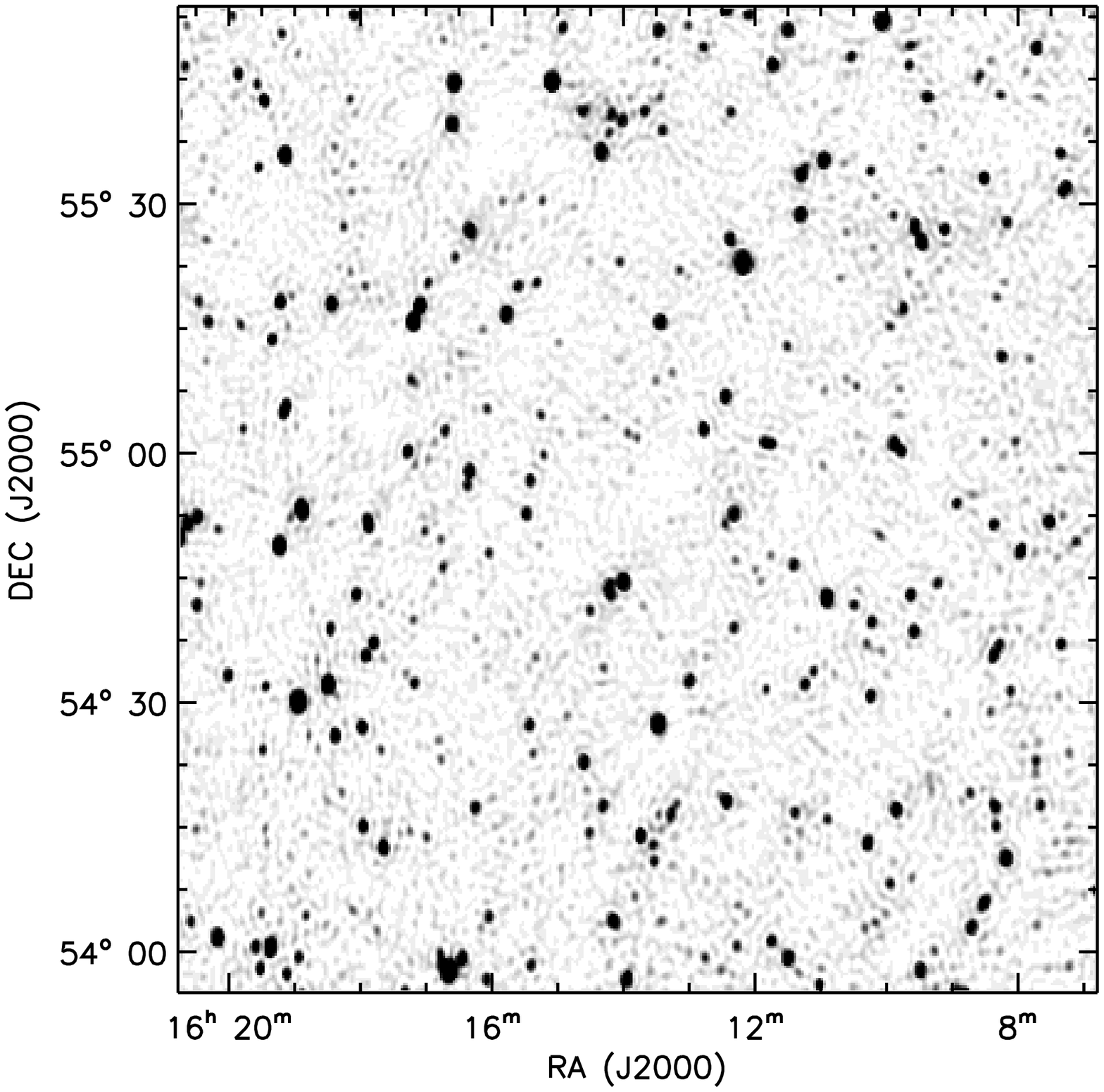}
\includegraphics[width=0.5\textwidth]{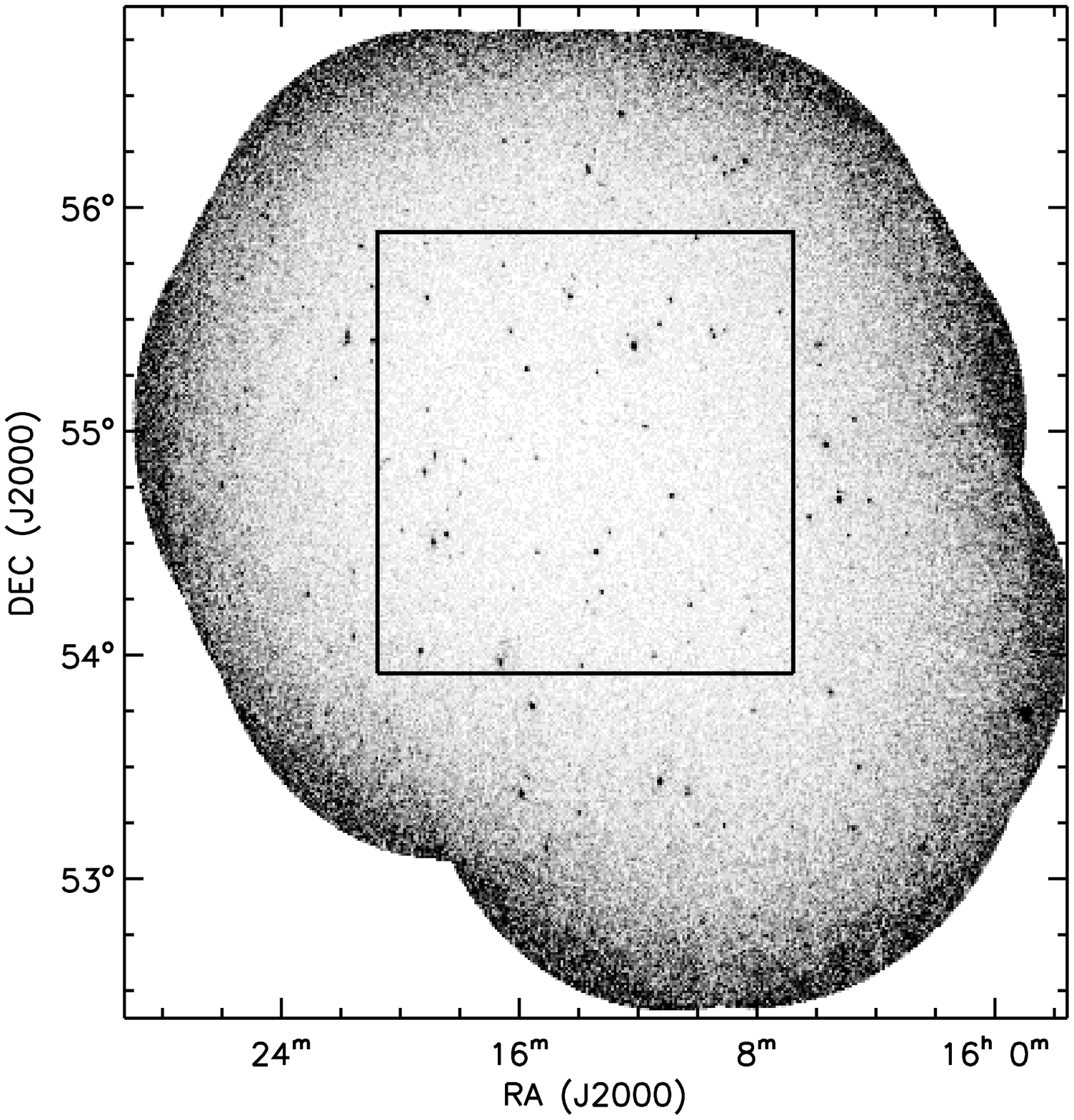}
\includegraphics[width=0.5\textwidth]{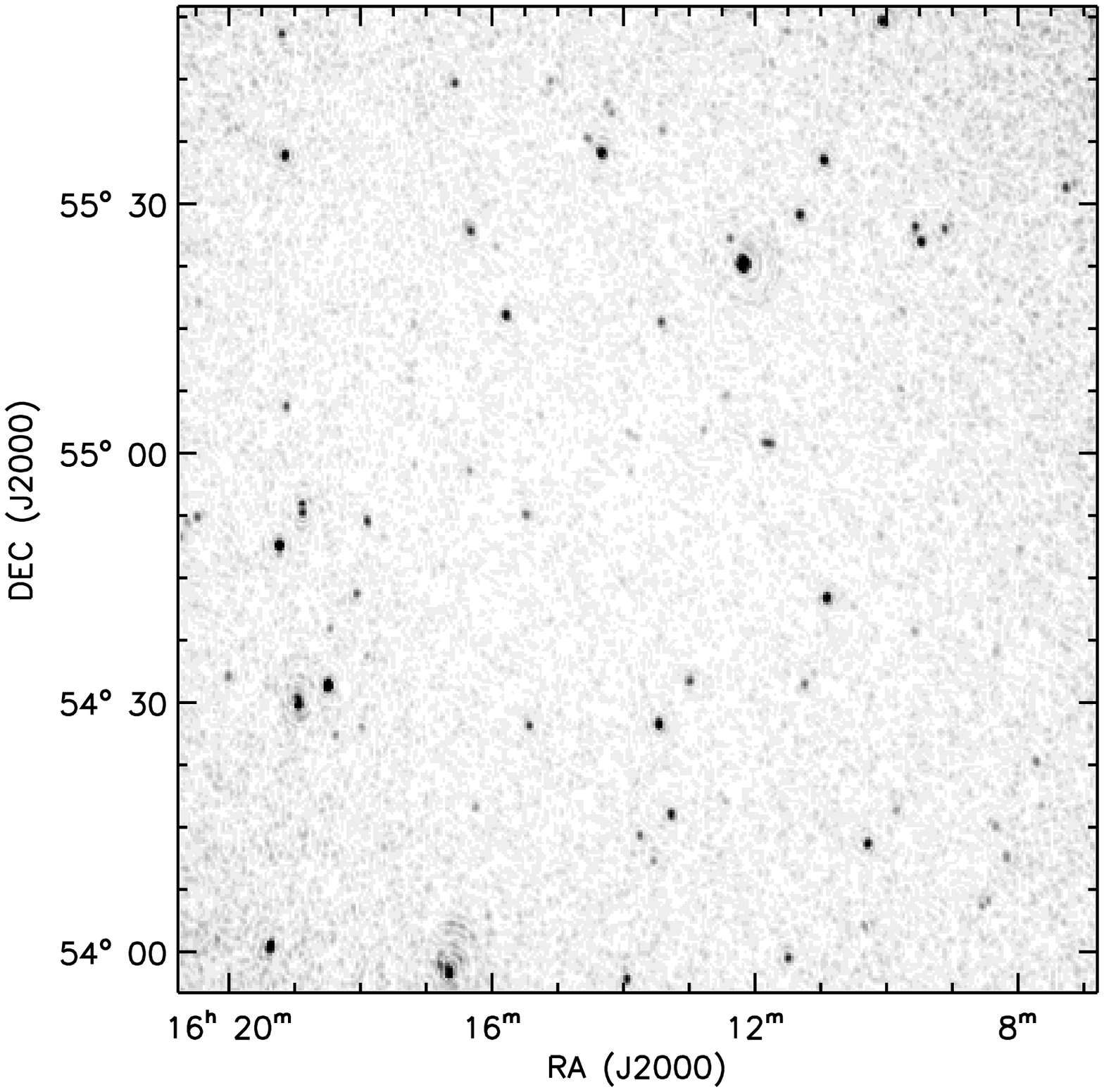}
\caption{The DRAO ELAIS N1 deep field observations.  \textit{Top panels:} Stokes $I$ with the linear greyscale set to the range $-0.1$ to $+1.5\,$mJy beam$^{-1}$.  \textit{Bottom panels:} Linear polarized intensity ($\sqrt{Q^2+U^2}$) with the linear greyscale set to the range $+0.0$ to $+1.0\,$mJy beam$^{-1}$.  The mosaics in the right-hand column are enlargements of the area outlined in the box of the images in the left-hand column.}\label{obsimg}
\end{figure}

The rms noise level was determined by fitting a Gaussian to the distribution of flux densities in the uniform noise mosaic (Figure \ref{noisegauss}).  The distribution of the pixel flux densities for Stokes $I$ is different from the distribution of Stokes $Q$ and $U$.  The negative and positive skewness in Stokes $I$ indicates a non-Gaussianity of the noise from confusion and detected sources.  The Stokes $Q$ and $U$ mosaics are well fitted by a Gaussian to flux densities of $150\,\mu$Jy ($\sim3\sigma$).  The non-Gaussian wings above the fit arise from real sources in the image and possibly also residual artifacts from cleaning the individual fields before mosaicing.  The central area of the mosaic has an rms noise level of $55\,\mu$Jy beam$^{-1}$ for Stokes $I$ and $45\,\mu$Jy beam$^{-1}$ for both Stokes $Q$ and $U$.  The Stokes $I$ and polarized intensity ($P = \sqrt{Q^2 + U^2}$) mosaics from the DRAO ELAIS N1 deep field can be seen in Figure \ref{obsimg}.

\section{Source Detection}
\subsection{Monte-Carlo Simulations}
The mosaics were searched for sources using an automated program that fits a two-dimensional Gaussian plus a constant level to each perspective source.  We ran Monte-Carlo simulations to (1) estimate and correct for any systematic fitting error, (2) estimate the probability of false detections and (3) estimate the source detection probability (completeness correction) as a function of flux density.  1000 simulations were run for each of total intensity and polarized intensity.  The simulated images had the same noise properties as the actual mosaics, since the source-subtracted mosaics were used as input noise mosaics for the simulations.  The source-subtracted mosaics were created by running the two-dimensional Gaussian fitting program on each of the total intensity and polarized intensity mosaics and removing the found sources with a fitted peak flux density greater than $4\sigma$ from the mosaic.  For each of the 1000 simulations, sources were seeded into the Stokes $I$ image with a flux density distribution following the fit of the 1.4 GHz source counts by \citet{Windhorst1990}.  The sources were input in random locations with an angular size equal to the beam making every input sources unresolved.  Since $\sim$10\% of Stokes $I$ sources were detected in polarization in the actual mosaics, 10\% of the simulated Stokes $I$ sources were selected as our input polarized sources.  Each source was then given a random $\Pi$ between 1 and 70\% and random polarization angle between $-90\arcdeg$ and $90\arcdeg$.  The polarized sources were then input into the Stokes $Q$ and $U$ images at the same location as their Stokes $I$ counterpart.

The statistical distribution of the noise in the polarized intensity image is non-Gaussian and has a non-zero mean.  The probability distribution of $P$ for sources with intrinsic polarized flux density of $P_{\rm 0}$ follows the Rice distribution \citep{Rice1945, Vinokur1965, Simmons1985},
\begin{equation}
f(P|P_{\rm 0}) = \frac{P}{\sigma}{\rm exp}\left[-\frac{(P^2 + P_{\rm 0}^2)}{2\sigma^2}\right]I_{\rm 0}\left(\frac{PP_{\rm 0}}{\sigma^2}\right),
\end{equation}
where $I_{\rm 0}$ is the modified Bessel function of the first kind.  If $P_{\rm 0}=0$ then the distribution reduces to the Rayleigh distribution,
\begin{equation}
f(P|0) = \frac{P}{\sigma}{\rm exp}\left(-\frac{P^2}{2\sigma^2}\right),
\end{equation}
which is the probability distribution of pixel amplitudes in the polarized intensity image in the absence of polarized emission.  The measured polarized flux density is corrected for the polarization noise bias following the approximation of \citet{Simmons1985}.   The estimate of $P_{\rm 0} = \sqrt{P^2 - \sigma^2}$ is a good approximation if the signal-to-noise ratio is larger than 4.

\begin{figure}
\epsscale{1.0}
\plotone{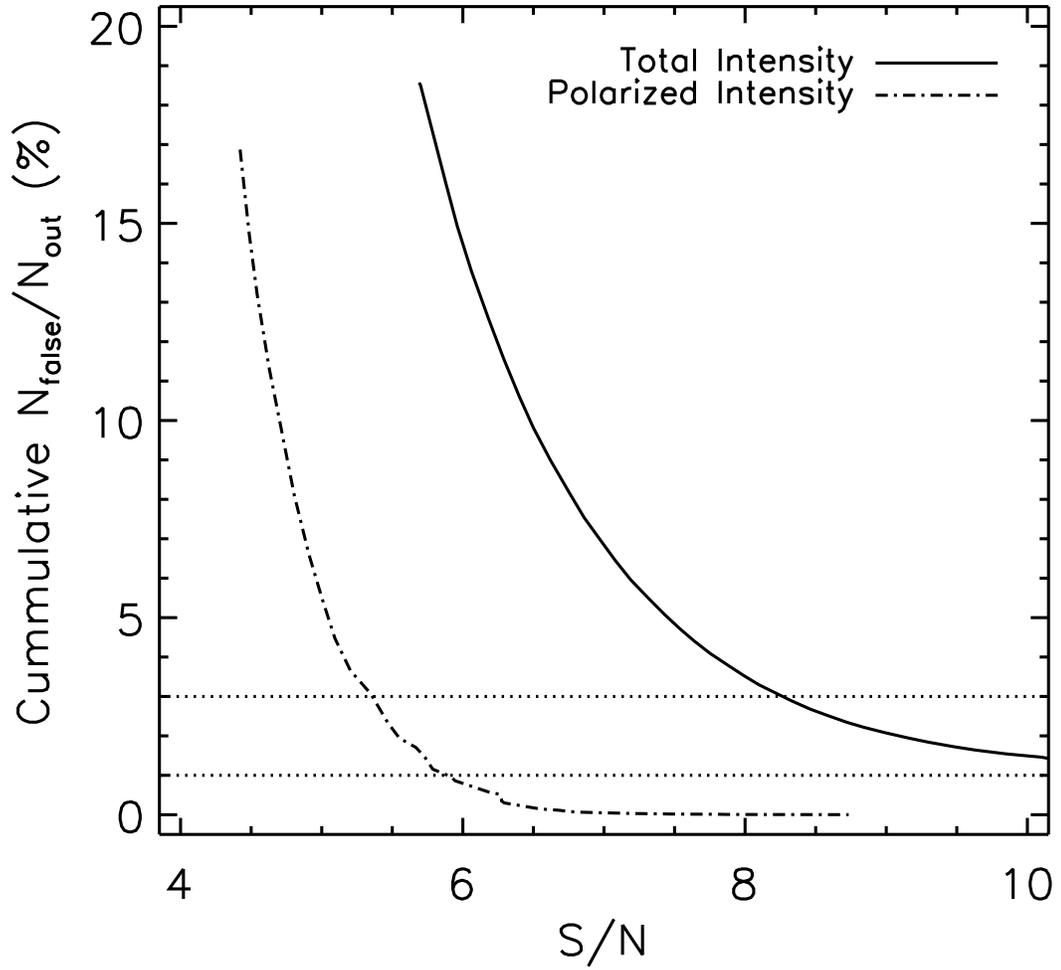}
\caption{The cummulative fraction of false detections as a function of signal-to-noise ratio.  The solid (---) and the dot-dashed (-- $\cdot$ --) curves represent the false detections in the total intensity and polarized intensity mosaics, respectively.  The horizontal dotted lines ($\cdot \cdot \cdot$) indicate the 3\% and 1\% false detection levels.}
\label{false}
\end{figure}

The results for the detected sources in the 1000 simulated images show that there are systematic biases in fitting both the peak flux density and angular size of the sources.  For the total intensity simulations, the peak flux density was over-estimated at a signal-to-noise ratio (SNR) of 15 at 1\% to 3\% at a SNR of 8.  The fitted angular size was over-estimated at a SNR of 15 at 1\% to 6\% at SNR of 8.  For polarized intensity simulations, the peak flux density and angular sizes  both show under-estimates of the input values.  The peak flux density was under-estimated at a SNR of 300 at 1\% to 15\% at SNR of 5.  The fitted angular size has the same value of under-estimates as the polarized peak flux density.  These were corrected for prior to any analysis.  The false detection rate from the 1000 simulated Stokes $I$ and polarized intensity images is shown in Figure \ref{false}.  The false detection rate reaches the 1\% level at SNR of 5.8 in polarization and 12.0 in total intensity.   The high SNR of false detections in total intensity is the result of confusion.  Since we require a polarized source to have a Stokes $I$ counterpart, the Stokes $I$ source list is cut at the 3\% level (SNR of 8) in order to not remove real polarized sources from our analysis.  The polarized source list is then cut at a SNR of 5.8.  From  the \citet{Taylor2007} polarized source list, 6 are not considered to be polarized sources in the new catalogue.  These 6 sources have a SNR below 5.8 confirming that a more conservative threshold for source detection was required.  Cutting the new source catalogue at a SNR of 8 in Stokes $I$ and 5.8 in polarization, we have allowed for accurate calculation of the completeness correction for the radio source counts and statistical analysis of the observed radio sources (see section 4.1).

\subsection{The DRAO ELAIS N1 Catalogue}
The sources in the actual mosaics were detected using the same procedure as in the simulations.  For sources that are located in both the total intensity and polarized intensity observational areas, there are 958 sources detected in the Stokes $I$ image down to SNR $\ge$ 8.  Of those, 136 have detectable polarization and the smallest fractional polarization detected of $0.73\,\pm\,0.12\%$.

The polarized sources are listed in Table \ref{catalog}.  The table gives the name, peak flux density $S$, \citet{Simmons1985} bias-corrected peak flux density $P_{\rm 0}$, percentage polarization $\Pi_{\rm 0}$, polarization angle PA $= 0.5 \,{\rm arctan}\,(U/Q)$ measured in the J2000 coordinate system, spectral index $\alpha_{1420-325}$ calculated between our flux density averaged over all four frequency bands and the WENSS source catalogue (see section 3.4), and FIRST classification (see section 3.3).

\subsection{Comparison with FIRST and NVSS}
\begin{figure}
\epsscale{1.0}
\includegraphics[width=0.5\textwidth]{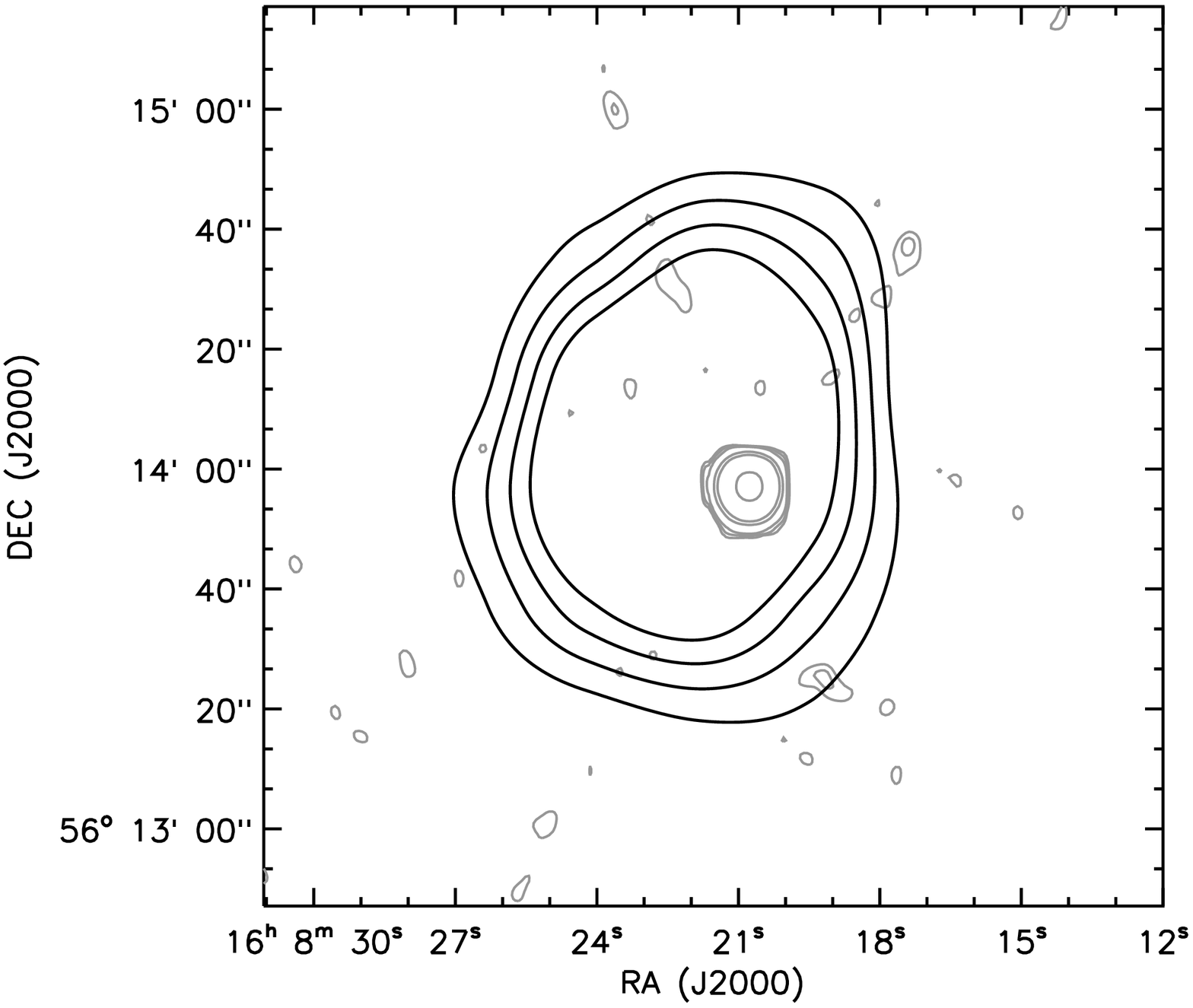}
\includegraphics[width=0.5\textwidth]{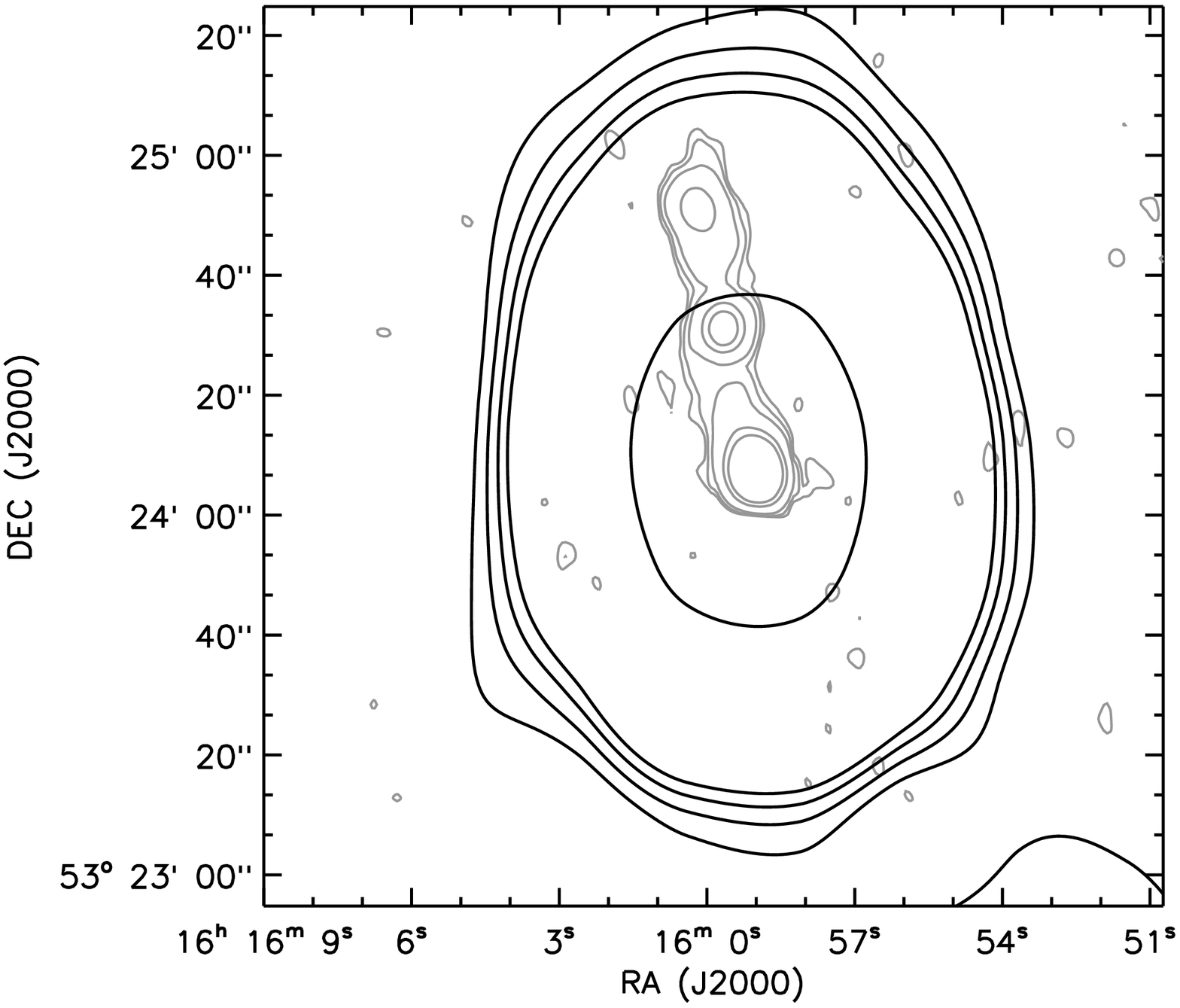}
\includegraphics[width=0.5\textwidth]{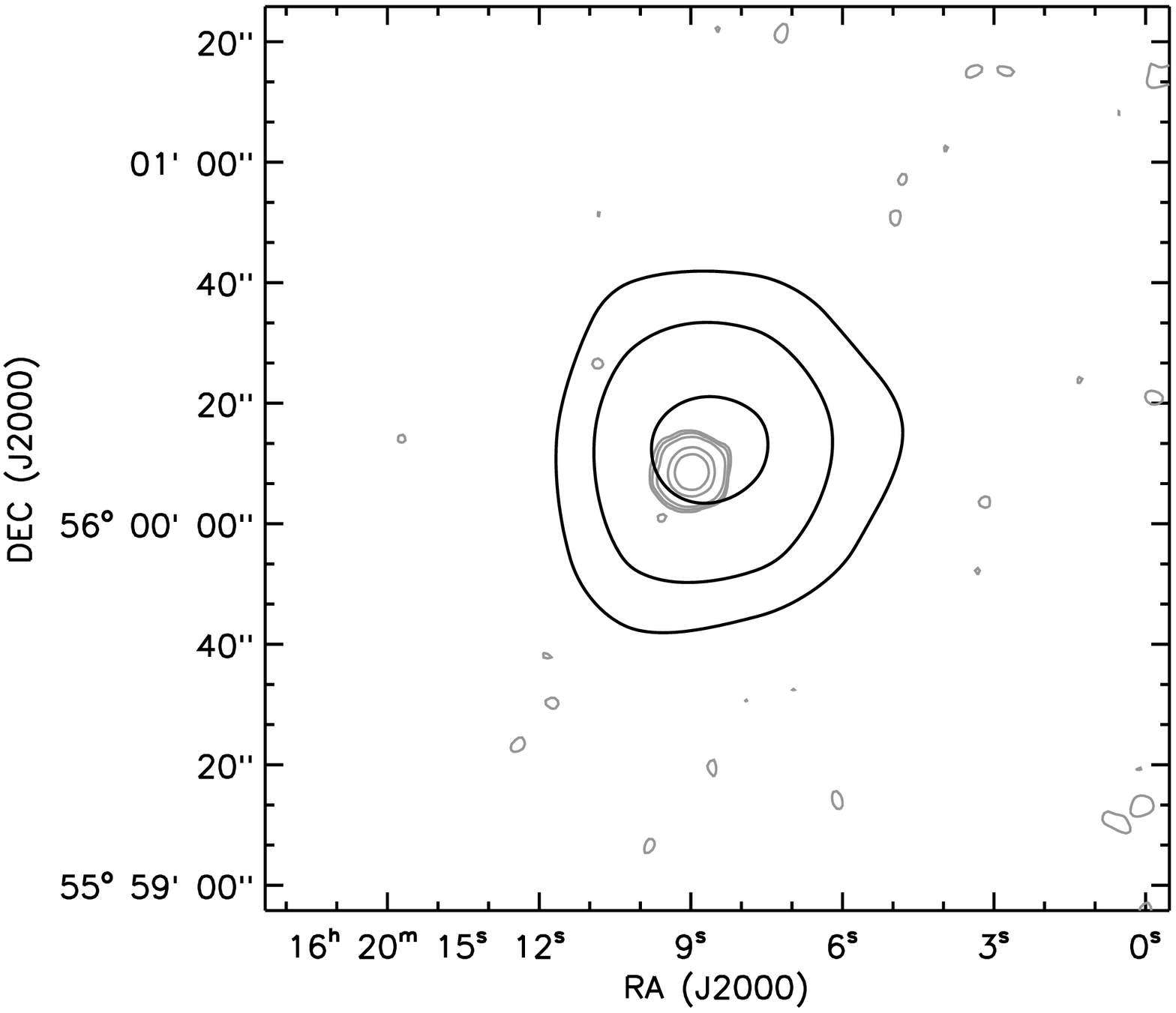}
\includegraphics[width=0.5\textwidth]{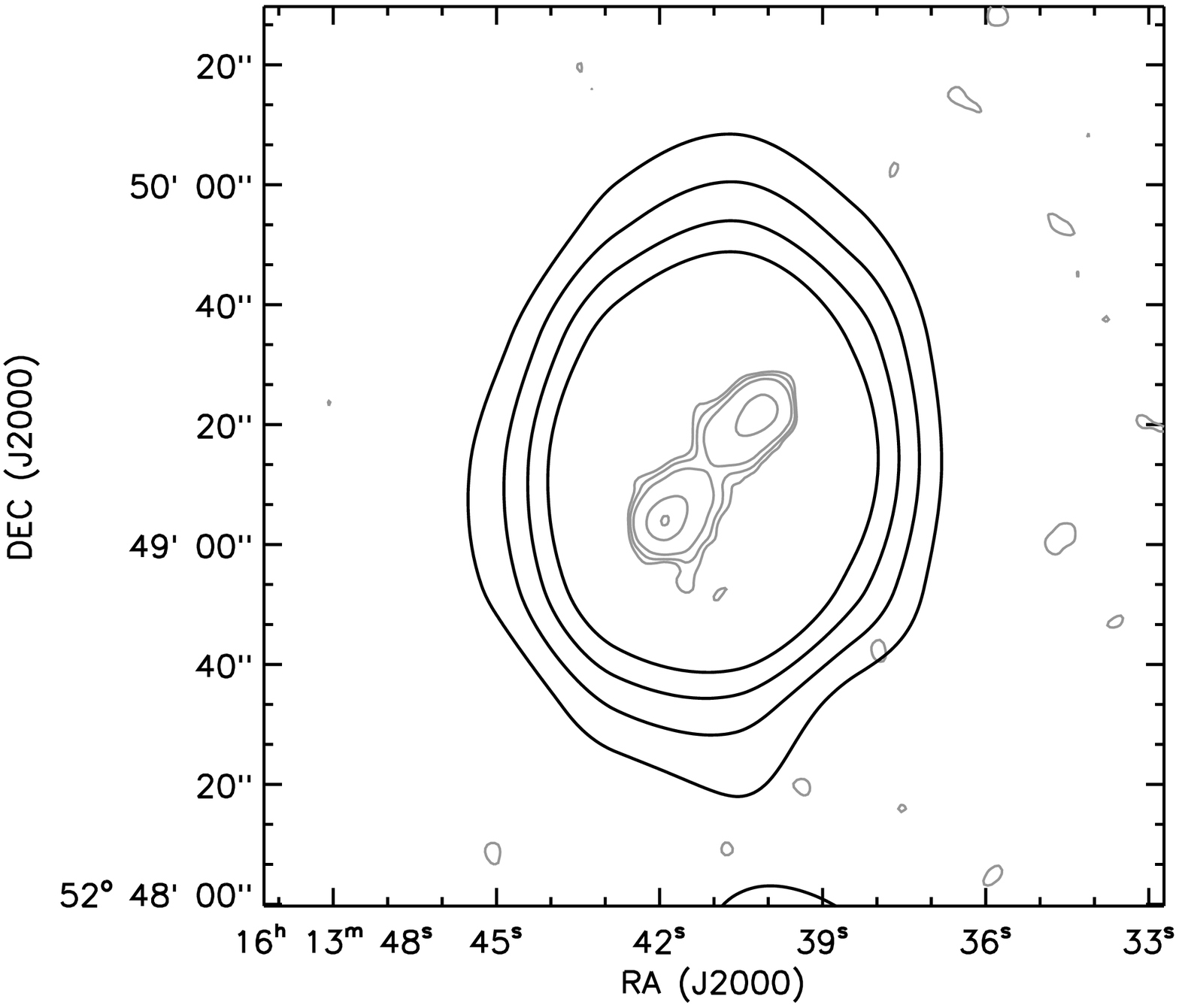}
\caption{Examples of the classification of FIRST counterparts: compact (left column), resolved (right column).  The black contours are the DRAO ELAIS N1 polarization at 4, 6, 8, 10, 50$\sigma_{\rm Q,U}$ and the grey contours are the FIRST total intensity at 3, 5, 10, 50, 100, 1000$\sigma_{\rm FIRST}$.}\label{firstsources}
\end{figure}

\begin{figure}
\epsscale{1.0}
\plotone{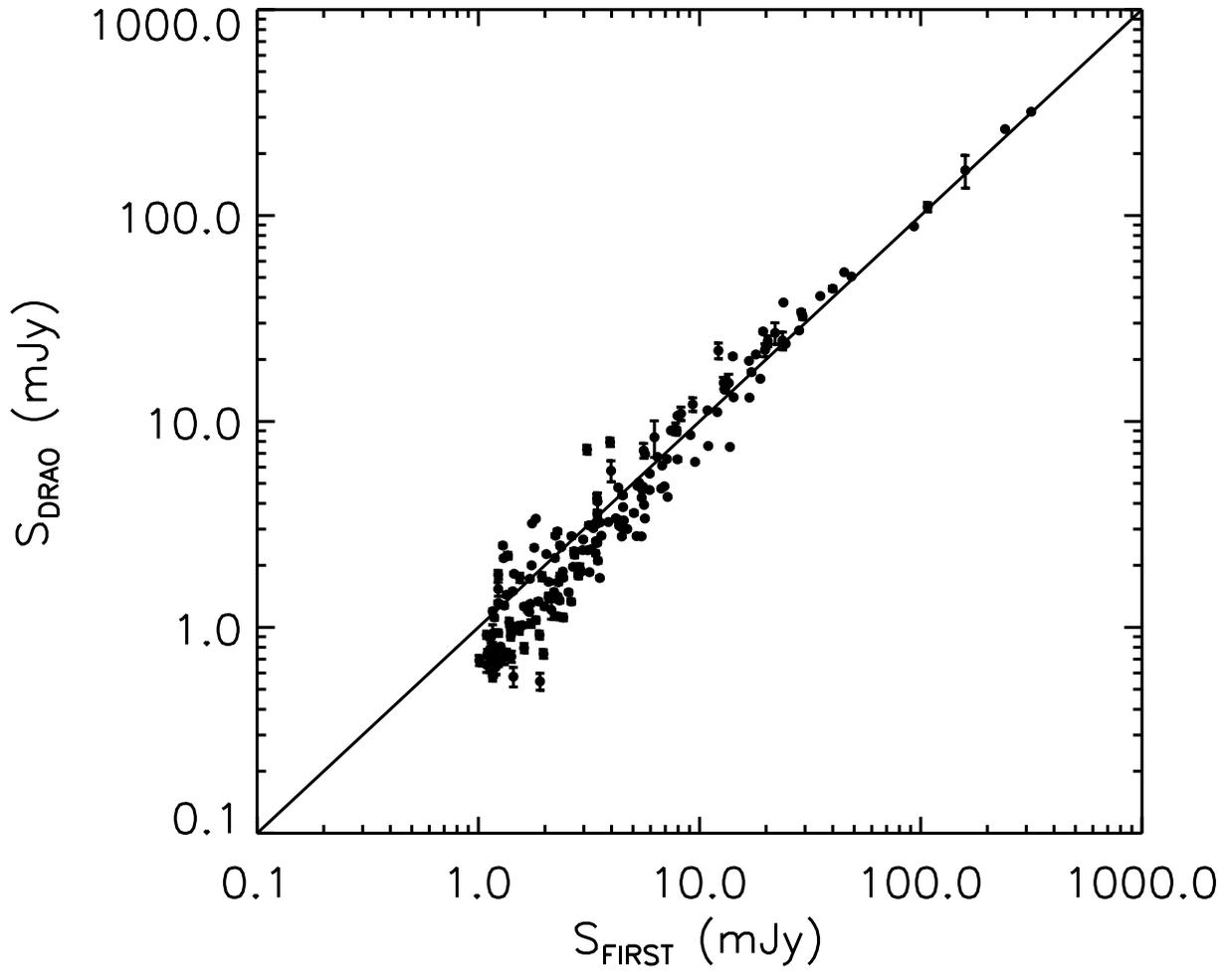}
\caption{The integrated total intensity flux density from DRAO compared to the integrated total intensity flux density from the corresponding FIRST source.  Only compact sources in FIRST were compared to their corresponding DRAO source.}\label{firstflux}
\end{figure}

\begin{figure}
\epsscale{1.0}
\plotone{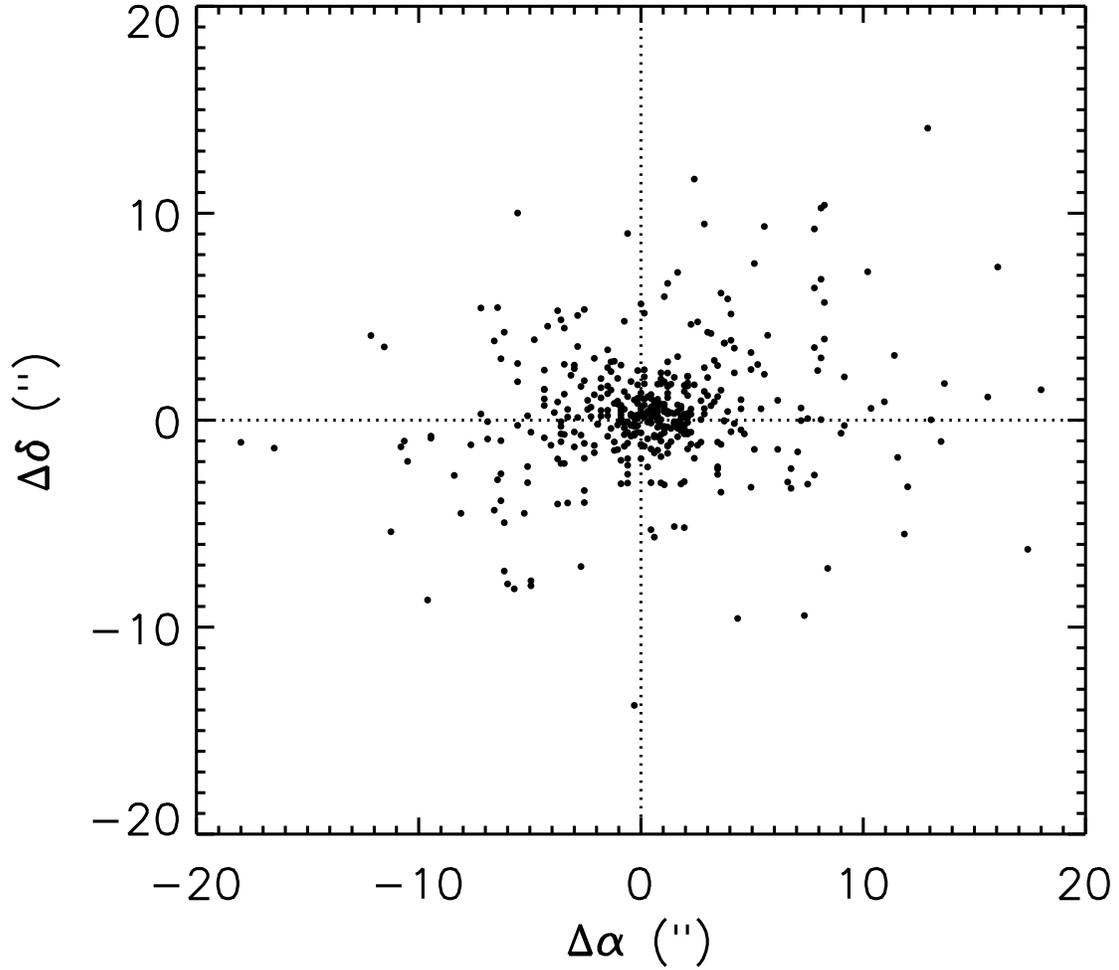}
\caption{The fitted DRAO source position compared to the corresponding FIRST source position.  Only compact sources in FIRST were compared to their corresponding DRAO source.  There is a systematic shift of the DRAO position compared to the FIRST position of $\Delta\alpha = 0.5 \pm 0.2\arcsec$ and $\Delta\delta = 0.4 \pm 0.2\arcsec$.}\label{firstpos}
\end{figure}

FIRST is a 1.4 GHz northern sky survey completed with the VLA in total intensity ($1\sigma=0.15\,$mJy) at an angular resolution of $5\arcsec$ \citep{White1997}.  The survey covers over 9,000 deg$^2$, including our DRAO ELAIS N1 deep field.  Of the 958 DRAO ELAIS N1 deep field sources, 795 have a DRAO flux density greater than 1.0 mJy, the flux density limit used for the FIRST source counts.  Of these, 697 have a match in the FIRST catalogue.  In some cases there are multiple FIRST sources within 1$\arcmin$ corresponding to one DRAO source.  These multiple FIRST sources are generally a single multi-component radio source, where the two lobes and central component are each clearly identified.  When this occurs, the DRAO source position is matched to the position of the central component, and classified as a resolved FIRST radio source.   A DRAO source is classified as a compact FIRST source in cases where there is only one compact FIRST source corresponding to one DRAO source.  See Figure \ref{firstsources} for examples of both classifications.  Another source matching situation occurs when a FIRST source is larger than $1\arcmin$ and resolved by the DRAO ST into two apparently independent sources, i.e two radio lobes separated by $>1\arcmin$.  In this case, 40 DRAO sources were found to correspond to 20 FIRST sources larger than $1\arcmin$.  In total, taking the multiple sources into account, there are 677 radio sources with $S \ge 1.0$ mJy in the DRAO ELAIS N1 deep field that are detected in FIRST.  Figure \ref{firstflux} shows a plot of DRAO flux density compared to FIRST flux density for sources classified as a compact FIRST radio source.  The position of each compact FIRST source was compared to the DRAO position (Figure \ref{firstpos}).  The fitted DRAO position agrees with the FIRST position to within 10$\arcsec$, half a pixel in the DRAO mosaic.  There is a systematic shift of the FIRST position compared to the DRAO position of $\Delta\alpha = 0.5 \pm 0.2\arcsec$ and $\Delta\delta = 0.4 \pm 0.2\arcsec$.  There are also 98 DRAO sources with a flux densities between $1.0<S<12.8\,$ mJy with no FIRST counterpart, well above the incompleteness cutoff of the FIRST survey. These sources may be resolved out in FIRST but not with the DRAO ST. 

The NVSS is a 1.4 GHz survey completed with the VLA in both total intensity and polarized intensity at an angular resolution of $45\arcsec$ \citep{Condon1998}.  The rms noise levels for the NVSS are $0.45\,$mJy beam$^{-1}$ in Stokes $I$ and $0.29\,$mJy beam$^{-1}$ in Stokes $Q$ and $U$.  There are $\sim 2 \times 10^6$ sources in the NVSS catalogue with $S \ge 2.0$ mJy.  NVSS counterparts were found for 597 out of a possible 613 sources with DRAO flux density $S \ge 2.0$ mJy.  The resolution of the NVSS is comparable to the resolution of DRAO and it is not surprising that in all cases there were single matches between the DRAO sources and the NVSS sources.  All of the 16 DRAO sources with no NVSS counterpart have flux densitites $\le4.1\,$mJy, and may point to the incompleteness of the NVSS in the 2$-$4$\,$mJy flux density range or these sources are variable.

\begin{figure}
\epsscale{1.0}
\plotone{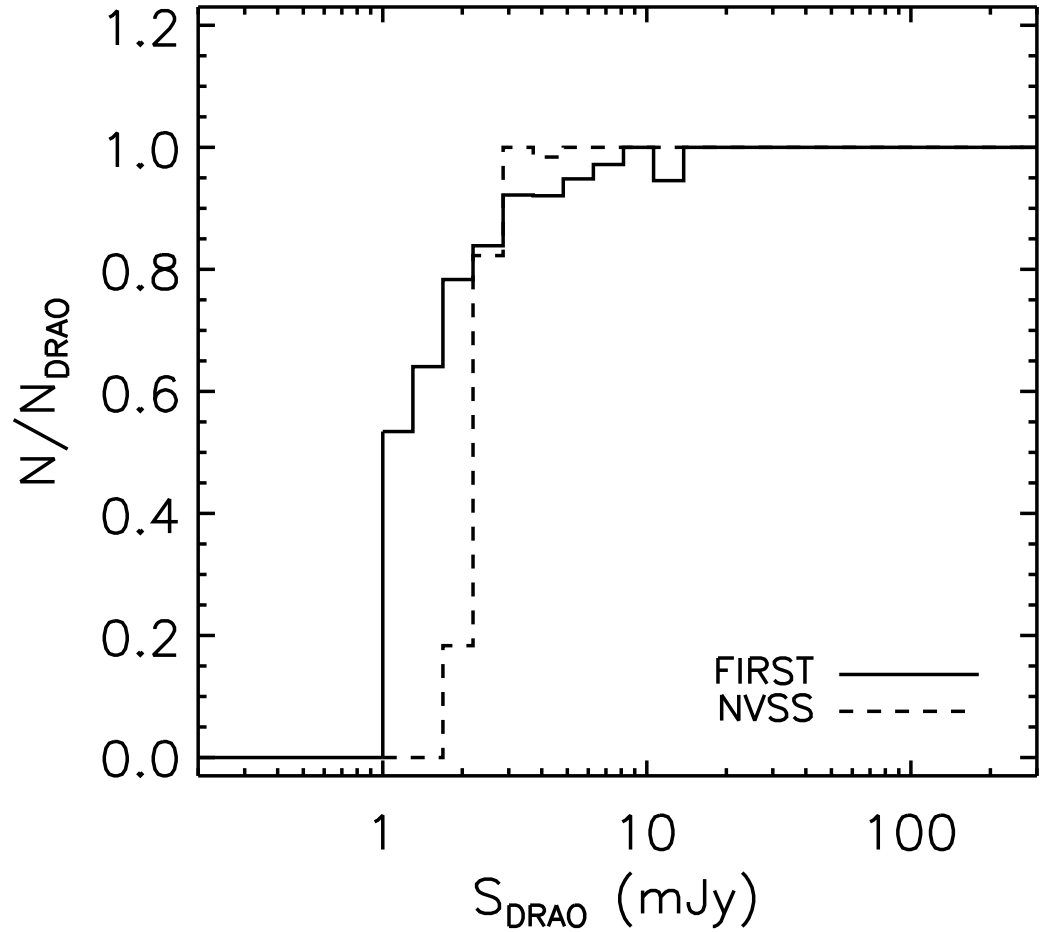}
\caption{The fraction of DRAO ELAIS N1 sources with matches in the FIRST (---) and NVSS (-- --) catalogues as a function of DRAO flux density.   The DRAO source that were found to have no corresponding FIRST object down to just above the NVSS flux density limit, have a positively detected NVSS source.}\label{detection}
\end{figure}

Figure \ref{detection} shows the fraction of DRAO sources matched in NVSS and FIRST as a function of DRAO flux density.  The DRAO sources are 100\% matched to a counterpart in both NVSS and FIRST down to $S_{\rm DRAO} = 12.8\,$mJy, at which point the fraction of sources with FIRST counterparts declines slowly to 50\% at the FIRST flux limit ($1\,$mJy).  The matching between DRAO and NVSS on the other hand remains at 100\% down to just above the NVSS flux density limit and then decreases rapidly to zero.  The DRAO sources that were found to have no corresponding FIRST object down to just above the NVSS flux density limit, have a positively detected NVSS source.  This difference of the NVSS and FIRST catalogues at $S \le 12$ mJy was also noted by \citet{Devries2002}, who matched sources for both surveys in the Bootes Deep Field.  The fact that there are sources not detected in FIRST (resolution of $5\arcsec$), but detected in both the DRAO ELAIS N1 deep field ($60\arcsec$ resolution) and the NVSS ($45\arcsec$ resolution), indicates that this is a resolution effect; i.e. diffuse emission is missed in FIRST.  Care must therefore be taken in statistical comparison of source data taken from observations with different angular resolutions.

\subsection{Comparison with WENSS}\label{wenss}
\begin{figure}
\epsscale{1.0}
\plotone{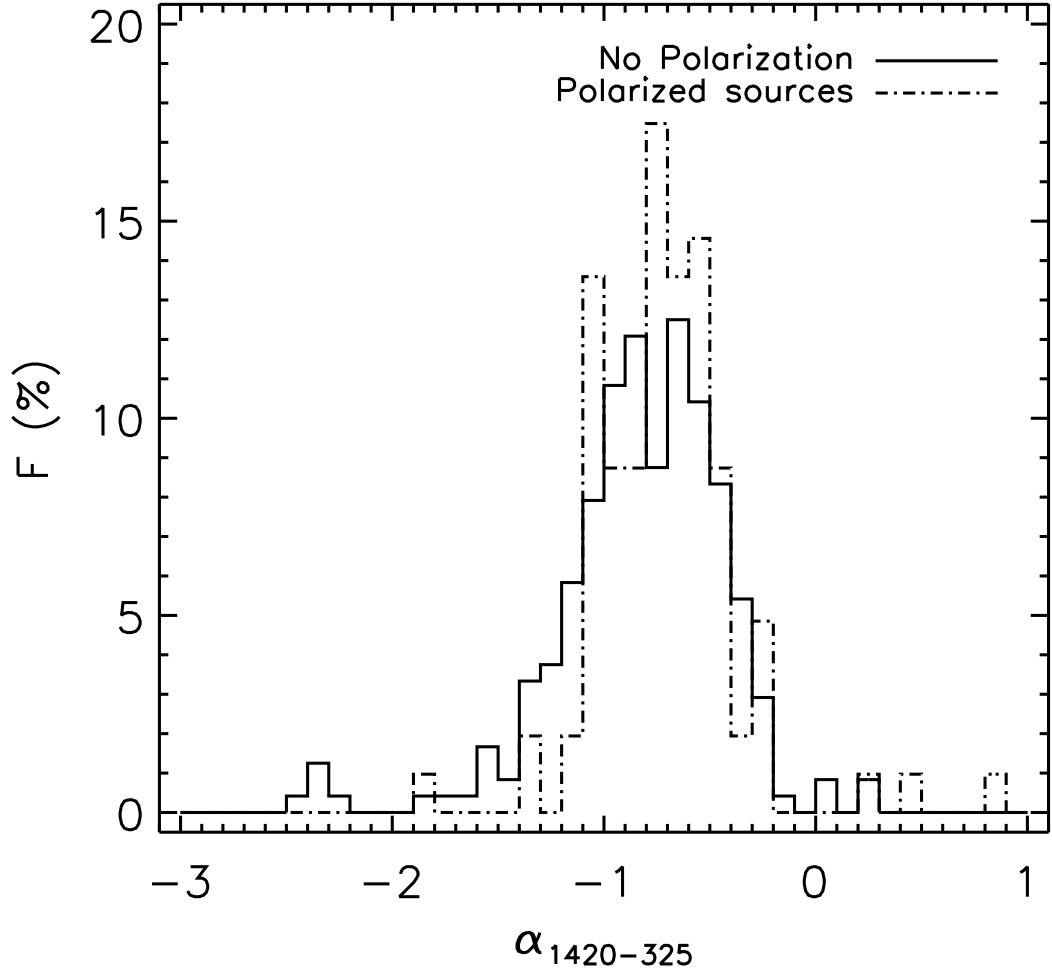}
\caption{DRAO (1.4 GHz) to WENSS (325 MHz) spectral index of 103 polarized sources (-- $\cdot$ --) and 240 sources with no detectable polarization (---).  The mean spectral index is $-0.71\,\pm\,0.04$ for polarized sources and $-0.82\,\pm\,0.03$ for sources with no detectable polarization.  The two distributions are no significantly different.}\label{spectralindex}
\end{figure}

WENSS is a northern sky survey at observed at 325 MHz and completed with the Westerbork synthesis telescope \citep{wenss}.  WENSS has an angular resolution of $54\arcsec \times 54\arcsec {\rm cosec}\delta$ and a limiting flux density level of $18\,$mJy ($5\sigma$).  343 DRAO sources have counterparts in the WENSS catalogue, including 103 of the 131 polarized sources.  The resolutions of the two surveys are comparable resulting in only single source matches between DRAO and WENSS sources.  The spectral index ($S \sim \nu^\alpha$) distribution of the 343 DRAO ELAIS N1 deep field sources is shown in Figure \ref{spectralindex}, for both polarized sources and sources with no detectable polarization.  The 103 polarized sources found in WENSS have a weighted mean of $\alpha_{1420-325} = -0.77\,\pm\,0.01$ and the 240 sources with no detectable polarization found in WENSS have a weighted mean of $\alpha_{1420-325} = -0.87\,\pm\,0.01$.  

\citet{DeBreuck2000} determined spectral indices for $143{,}000$ radio sources between NVSS and WENSS and showed that this radio source population consists of a steep-spectrum galaxy and flat-spectrum quasar part separated by a spectral index of about $-0.3$.  Following this distinction between steep- and flat-spectrum sources, we have 95 steep-spectrum polarized sources and 8 flat-spectrum polarized sources.   For sources with no detectable polarization we have 228 steep-spectrum sources and 12 flat-spectrum sources.  The fraction of steep-spectrum to flat-spectrum sources shows little difference between polarized and unpolarized sources.  The hypothesis that the distribution of polarized sources and the sources with no detectable polarization in Figure \ref{spectralindex} were drawn from the same spectral index distribution was rejected by the Kolmogorov-Smirnov test at only the 94\% confidence level.

\begin{figure}
\epsscale{1.0}
\plotone{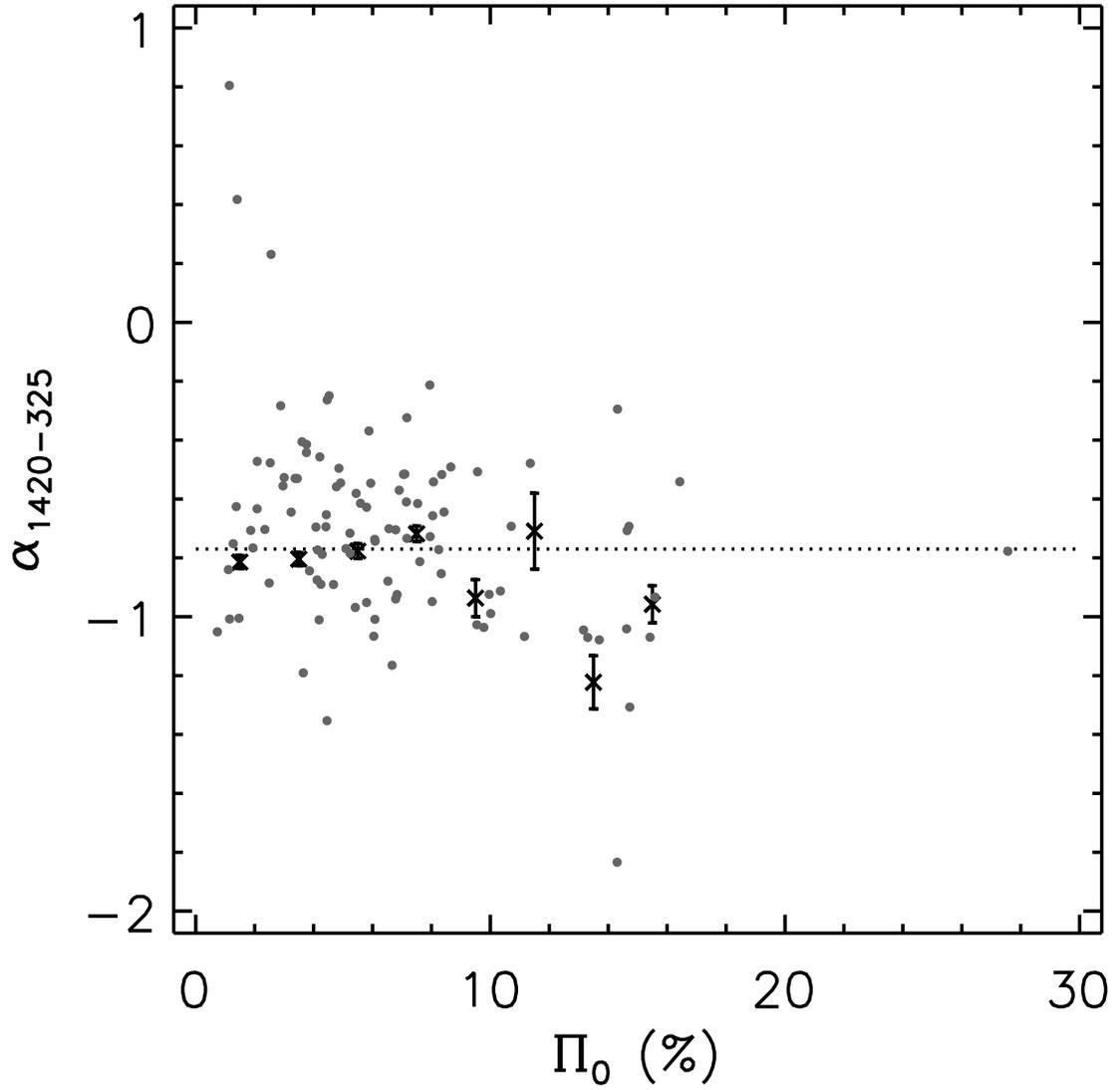}
\caption{Spectral index of DRAO ELAIS N1 polarized sources as a function of fractional polarization $\Pi_{\rm 0}$.  The $\times$ symbols indicate the weighted average of the spectral index in fractional polarization bins. The dotted line ($\cdot$ $\cdot$ $\cdot$) indicates the mean spectral index of the polarized sources.}\label{alphapi}
\end{figure}

Figure \ref{alphapi} shows a plot of the spectral index as a function of the degree of polarization for the 103 polarized DRAO sources with a WENSS counterpart.  The horizontal line indicates the mean spectral index for polarized sources.  Calculating the weighted average in bins of percentage polarization shows that the spectral index remains almost constant around the mean spectral index of $-0.77$ for polarized radio sources with $\Pi_{\rm 0} < 11$\%.  \citet{Gardner1969} noted that for radio galaxies with $20\,$cm flux densities above $2.5\,$Jy there is a weak trend of an increase in fractional polarization with a decrease in spectral index.  However, they found no trend with objects observed at 11 cm.  On the other hand \citet{Kronberg1970} note that there is an opposite correlation between fractional polarization and spectral index for bright sources observed at 49 cm.

\section{Nature of Polarized Radio Sources}
\subsection{Source Counts}
\begin{figure}
\epsscale{1.0}
\plotone{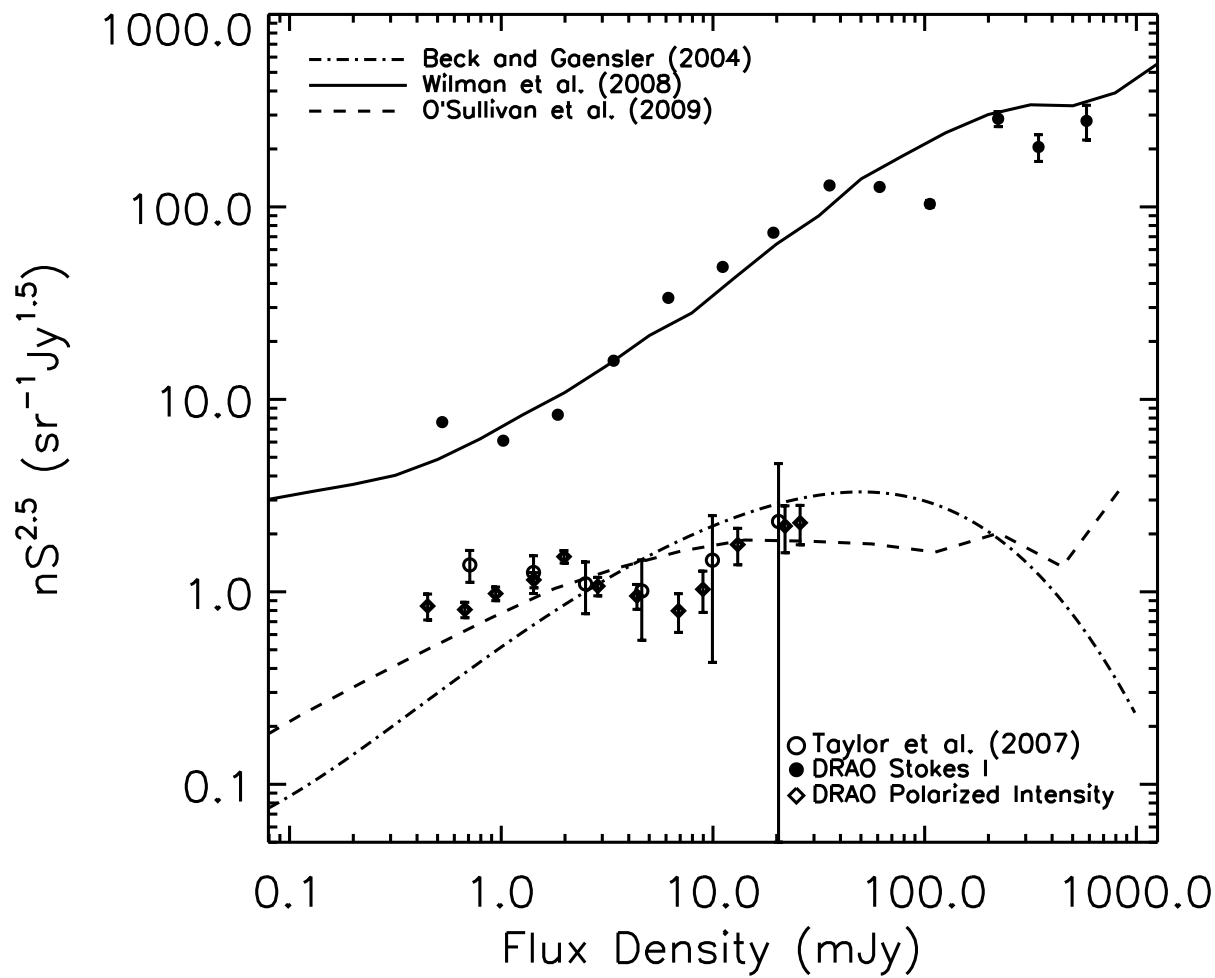}
\caption{Euclidean-normalized differential source counts for total intensity (closed circles) and polarized intensity (diamonds), along with polarized radio source counts from \citet{Taylor2007} (open circles).  The solid line is the \citet{Wilman2008} model of total intensity source counts.  The dashed line is the polarized source count model from \cite{OSul2008}, while the dot-dahsed line is the polarized source counts derived by \citet{Beck2004}. }
\label{counts}
\end{figure}

The Euclidian-normalized differential source counts of the DRAO ELAIS N1 deep field are shown in Figure \ref{counts} for both total intensity and polarized intensity. The Stokes $I$ source counts are plotted down to a flux density of $500\,\mu$Jy and the polarized source counts (Table \ref{polcounts}) are plotted down to $400\,\mu$Jy.  The polarized source counts do not include the fainteset 8 polarized sources as the error on the count does not provide an accurate representation of the polarized source counts below $400\,\mu$Jy.  The total intensity source counts follow the \citet{Wilman2008} semi-empirical simulation of the total intensity sky (solid curve in Figure \ref{counts}).  Our polarized source counts can be compared to those of \citet{Taylor2007} as well as those derived by \citet{Beck2004} and \citet{OSul2008}, each of which is plotted in Figure \ref{counts}.  We confirm the \citet{Taylor2007} result that sources with $P_{\rm 0} < 3$ mJy are found in significantly greater number than what is predicted from the bright source population.  The drop in the polarized source counts around polarized flux density range of $3-10\,$mJy is the result of total intensity sources in the total intensity flux range of $50-100\,$mJy, which is shown to drop as well.  The local surface density non-uniformity of sources in these flux density ranges is most likely the cause for the observed drop in the radio source counts.  We can also compare our polarized radio source counts with the \citet{OSul2008} model, which is based on the total intensity sky of \citet{Wilman2008} and a polarization distribution derived from a luminosity dependence for the polarization of FRI and FRII radio galaxies.  Again, our results lie above the predictions indicating potentially a strong dominant dependence on fractional polarization.  

\begin{figure}
\epsscale{1.0}
\plotone{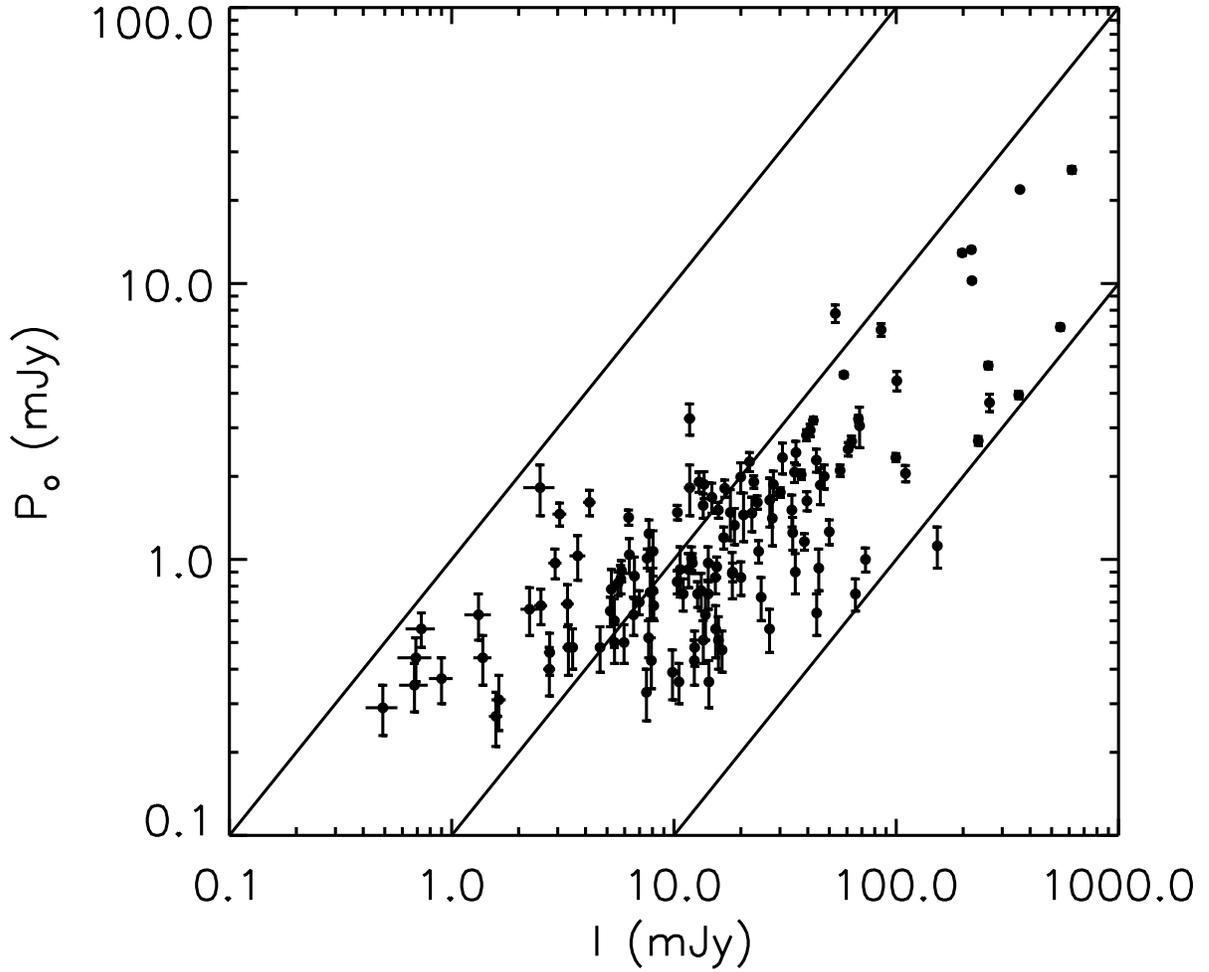}
\caption{Distribution of log$I - $ log$P_{\rm 0}$ of observed polarized sources.  The solid diagonal lines indicate the $\Pi_{\rm 0}$ = 1\% (right), 10\% (middle), and 100\% (left) fractional polarization levels.}\label{logilogp}
\end{figure}

Information of the nature of polarized radio sources can be determined from the fractional polarization $\Pi_{\rm 0}$. A comparison of $I$ and $P_{\rm 0}$ for the polarized sources in the DRAO ELAIS N1 deep field can be seen in Figure \ref{logilogp}.  The solid lines indicate the loci of sources with $\Pi_{\rm 0} = 100\%, 10\%$ and $1\%$.  There is a clear increase in the number of sources with higher fractional polarization with a decrease in total flux density.  

\begin{figure}
\epsscale{1.0}
\plotone{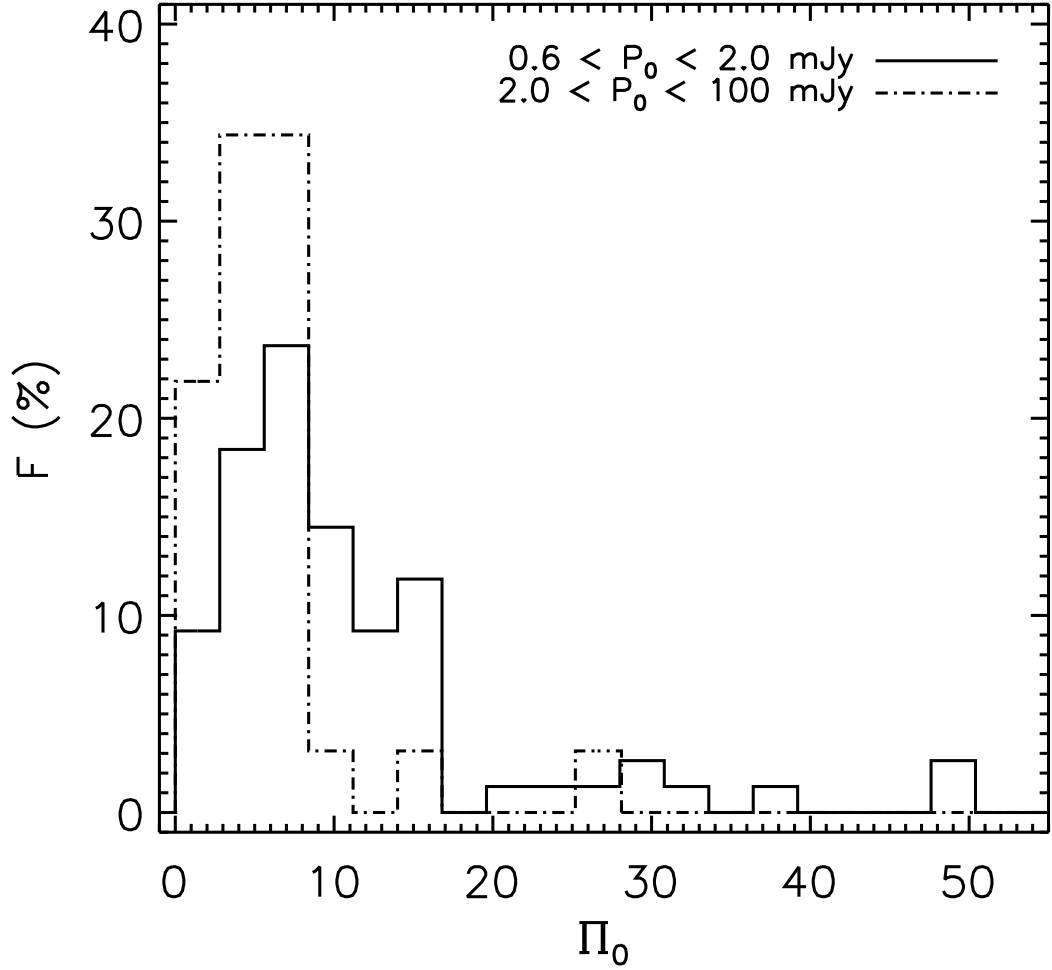}
\caption{Histogram of the fraction having different percentage polarization for the 37 sources in the polarized intensity range of $0.6 < P_{\rm 0} < 2.0$ mJy (---) and the 76 sources in the polarized intensity range of $2.0 < P_{\rm 0} < 100$ mJy (-- $\cdot$ --).  The flux density ranges are taken from Figure \ref{logilogp} to show the difference in fractional polarization as flux density decreases.}\label{logilogphist}
\end{figure}

The change in the distribution of $\Pi$ with flux density is illustrated in Figure \ref{logilogphist}.  Two different polarized flux density ranges are shown: $0.6 < P_{\rm 0} < 2.0$ and $2.0 < P_{\rm 0} < 100\,$mJy.  Selecting flux density bins in $P_{\rm 0}$ instead of $S$ allows us to be complete in our polarized source sample, and the high signal-to-noise of the bins allows us to be sure that the sources are not effected by noise bias.  Selecting flux density bins in $P_{\rm 0}$ instead of $S$ allows us to be complete in our polarized source sample, and the high signal-to-noise of the bins allows us to be sure that the sources are not effected by noise bias.  For a complete sample of sources, the distribution in $\Pi_{\rm 0}$ should be the same in both $P_{\rm 0}$ flux bins if the distribution of $\Pi_{\rm 0}$ does not change.  The same selection is imposed on the data in both bins, leaving only the fact that the upper bin contains fewer sources.  There is a higher fraction of polarized sources in the lower polarized flux density bin with $\Pi_{\rm 0} > 10$ \%, than in the higher polarized flux density bin. The hypothesis that the two polarized flux density distributions were drawn from the same distribution was rejected by the Kolmogorov-Smirnov test at the 99\% confidence level.  

Unfortunately, the DRAO polarization observations are not yet sensitive enough to detect the polarization emission from starburst and normal galaxies.  Since there is no polarized emission from these galaxies and the polarized source counts lie slightly above the \citet{OSul2008} model, we conclude that the polarized radio sources with $P_{\rm 0} < 3$ mJy are more highly polarized then the strong source population. There must be a dependence on flux density, fractional polarization and luminosity that plays a stronger role at the fainter levels of the polarized source counts.  Future observations will hopefully allow us to complete source counts for the intrinsically faintest population of galaxies.

\subsection{Structure of Polarized Radio Sources}
The structure of polarized radio emission can provide information on the intrinsic properties of the source.  The mean spectral index of the DRAO polarized sources indicates that the majority of these polarized objects are steep-spectrum objects, which contain both diffuse and compact emission components.  However, the resolution of the DRAO ELAIS N1 deep field data does not provide enough information to determine if the polarized emission is originating from the diffuse or compact components.  The faintest Stokes $I$ flux density of a DRAO source with a WENSS counterpart is 1.19 mJy.  There are 751 DRAO sources down to a DRAO flux density of $1.19\,$mJy, 131 of which are polarized.  343 of the 751 sources are matched with a WENSS counterpart corresponding to $46 \pm 3\%$ of our sample. Table \ref{wensstable} shows that there is $79 \pm 8\%$ polarized sources are matched in WENSS while just $39 \pm 3\%$ of source with no detectable polarization are matched.

A comparison of the 677 FIRST sources in our deep field show that there are 374 FIRST objects ($55 \pm 3\%$ of our sample) with no visible structure making them compact radio sources.  The other 303 FIRST objects ($45 \pm 3\%$ of our sample) were resolved.  115 of the 677 DRAO sources with FIRST counterparts have detectable polarization.  As outlined in Table \ref{firstpercent}, $23 \pm 5\%$ of the polarized sources with a FIRST counterpart are classified as unresolved or compact objects whereas $77 \pm 8\%$  are considered to be resolved.  For the sources with no detectable polarization the statistics are almost reversed, $62 \pm 3\%$ are compact or unresolved and $38 \pm 3\%$ are classified as resolved objects.  This indicates that polarized sources tend to have structure at arc-second scales and that the polarized emission is most likely not beamed.  This confirms that the polarized radio sources tend to be lobe-dominated radio galaxies.

\begin{figure}
\epsscale{1.0}
\includegraphics[width=0.5\textwidth]{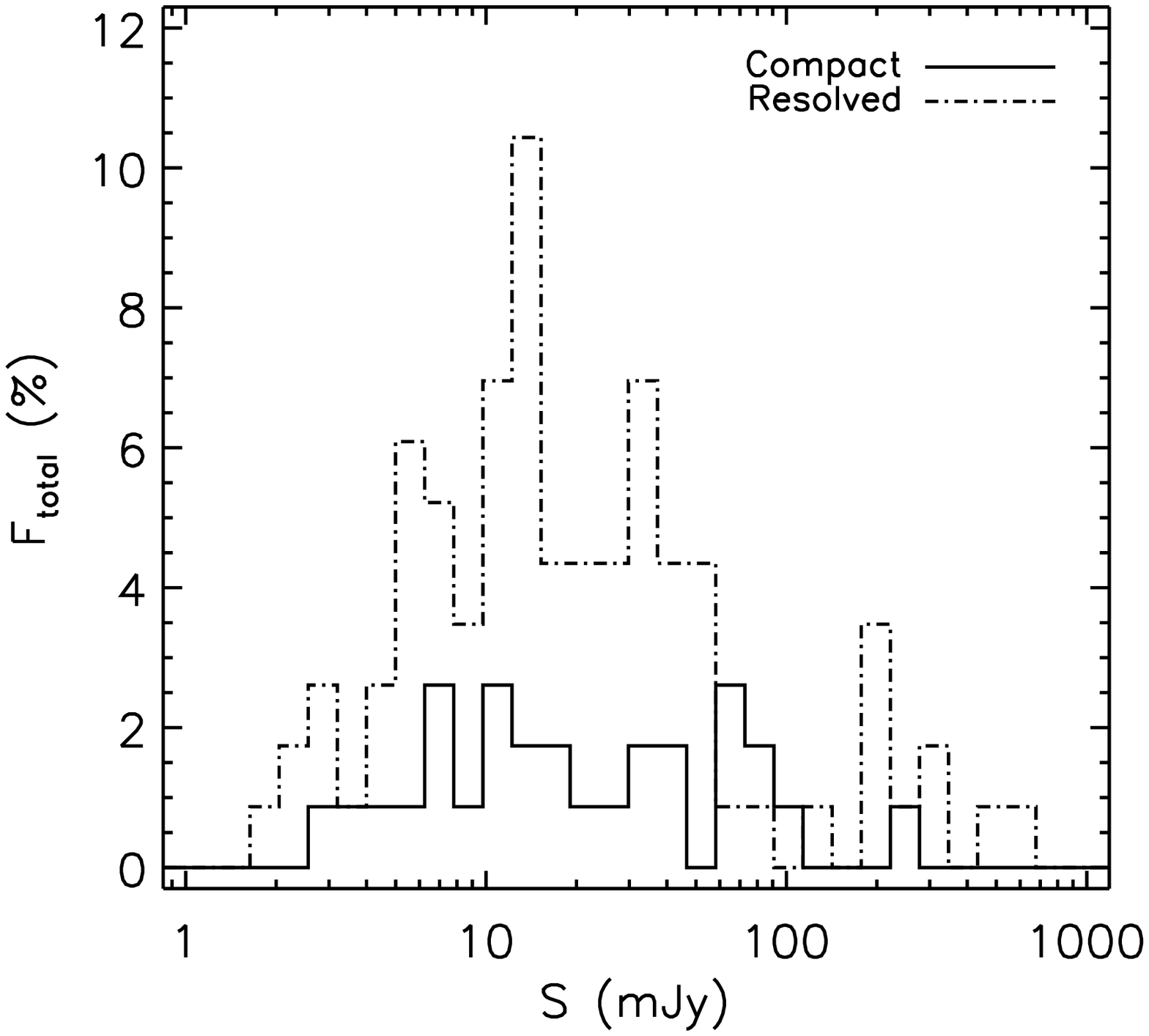}
\includegraphics[width=0.5\textwidth]{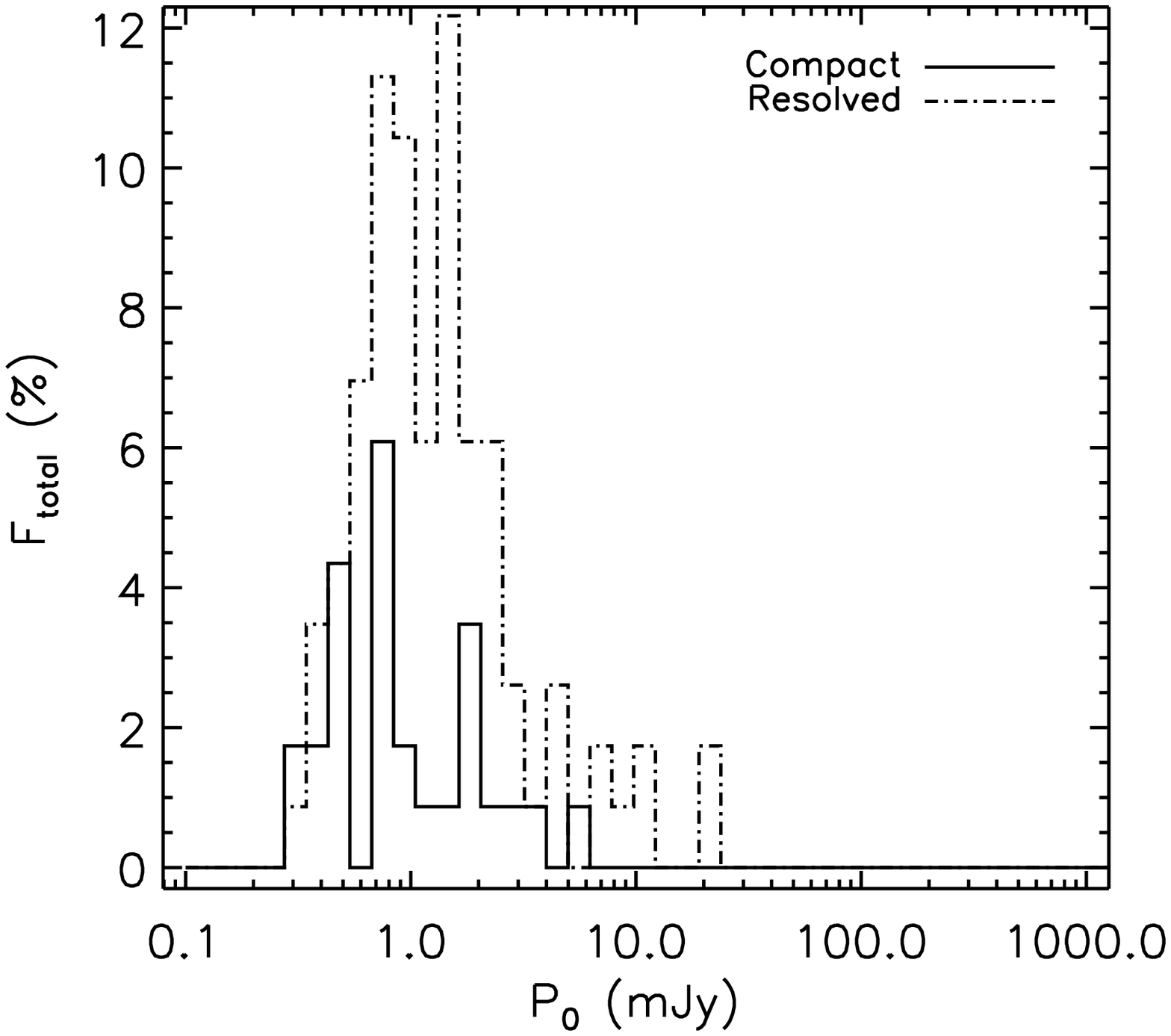}
\caption{Histograms of the total number of polarized sources classified as compact (---) or resolved (-- $\cdot$ --) in FIRST.  \textit{Left:} The distribution of polarized sources as a function of Stokes $I$ flux density.  \textit{Right:} The distribution of polarized sources as a function of polarized intensity.}\label{piclass}
\end{figure}

Figure \ref{piclass} shows a histogram of the DRAO-FIRST matched polarized as either resolved and compact source as a function of total intensity and polarized intensity.  There is a larger number of resolved sources compared to compact sources at fainter polarized intensity levels.  This increase in the number of resolved polarized sources occurs for $P_{\rm 0} < 3$ mJy and $S < 60$ mJy.   Resolved polarized sources are likely responsible for the increase in the polarized source counts at $P_{\rm 0} < 3\,$mJy, as seen in Figure \ref{counts}.  The median fractional polarization of the sources classified as resolved is 6.8 $\pm$ 0.7\% whereas the median fractional polarization of the compact sources is 4.4 $\pm$ 1.1\%.  The DRAO ELAIS N1 deep field polarized sources classified as resolved in FIRST are more polarized than the compact sources.  \citet{Hogbom1974} observed 79 sources with $S > 1.0\,$Jy in polarization with the Westerbork synthesis radio telescope at 1.4 GHz and noticed that extended well resolved diffuse areas have higher degree of polarization than the compact source components.  Our results confirm that this trend continues for polarized sources down to $P_{\rm 0} \ge 270\,\mu$Jy.  This indicates that the higher degree of polarization may be originating in the lobe dominated structure.  However, further high resolution polarimetric observations will be able to determine the magnetic field structure of these highly polarized resolved radio sources.

\section{Conclusion}
We present the final observations of the DRAO Planck Deep Field in the ELAIS N1 region along with a complete sample of compact polarized radio sources.  The 40 field observations at 1.4 GHz reach a sensitivity of $55\,\mu$Jy beam$^{-1}$ in Stokes $I$ and $45\,\mu$Jy beam$^{-1}$ in Stokes $Q$ and $U$.  There are 958 robustly detected radio sources in the 15.16 deg$^2$ of the observations down to a signal-to-noise of 8, 136 of which have detectable polarization.

The polarized source counts from the DRAO ELAIS N1 deep field are presented down to a polarized intensity of $400\,\mu$Jy.  The polarized source counts below $P_{\rm 0} < 3\,$mJy are higher than published polarized radio source count models.  This increase shows a continuing trend of increased polarization fraction with decreasing flux density, confirming the \citet{Taylor2007} result.  We detect starburst and normal galaxies in the total intensity observations, but we have yet to reach the sensitivity required to detect the polarized emission from these galaxies.  Our complete polarized source sample consists primarily of FRI and FRII radio galaxies.  

The DRAO deep field sources were matched to counterparts in the FIRST, NVSS and WENSS catalogues.  We found that a higher fraction of polarized radio sources are detected in WENSS than those with no detectable polarization.   The mean spectral index of the polarized sources is $-0.77\,\pm\,0.01$ compared with $-0.87\,\pm\,0.01$ for sources with no detectable polarization.  The majority of polarized sources are steep spectrum objects.  Moreover, we find no correlation between spectral index and fractional polarization.

The DRAO deep field sources with a FIRST counterpart  show that a higher fraction of polarized sources show structure at arc-second scales than the sources with no detectable polarization.   There is a larger fraction of resolved polarized sources, 77 $\pm$ 8\%, than for sources with no detectable polarization, 38 $\pm$ 3\%.   The median polarization of sources classified as compact is 4.4 $\pm$ 1.1\% and 6.8 $\pm$ 0.7\% for polarized sources classified as resolved objects in FIRST.  One possible explanation for the increase in the polarized source counts at $P_{\rm 0} < 3\,$mJy is that the resolved sources are the dominant objects at these flux density levels.  However, there is currently no information on the origin and location of the polarized emission of these faint polarized radio galaxies.  

\acknowledgments
Observations and research on the DRAO \textit{Planck} Deep Fields are supported by the Natural Science and Engineering Council of Canada and the National Research Council Canada.  The Dominion Radio Astrophysical Observatory is operated as a National Facility by the National Research Council of Canada.


\clearpage




\begin{deluxetable}{lrrrrrrcc}
\rotate
\tablewidth{0pt}
\tabletypesize{\footnotesize}
\tablecaption{Polarized sources from the DRAO ELAIS N1 source catalog.}
\tablehead{
\colhead{Source Name} & \colhead{RA (J2000)} & \colhead{DEC (J2000)} & \colhead{$S$} & \colhead{$P_{\rm 0}$} & \colhead{$\Pi_{\rm 0}$} & \colhead{$PA$} & \colhead{$\alpha_{1420-325}$} & Class.\tablenotemark{*}\\
\colhead{ } & \colhead{ } & \colhead{ } & \colhead{(mJy)} & \colhead{(mJy)} & \colhead{(\%)} & \colhead{($\arcdeg$)} & \colhead{ } &\colhead{}
}
\startdata
DRAO\_J160016.66+533944.82 & 16 00 16.66 $\pm$  0.91 & 53 39 44.82 $\pm$  0.28 & 616.34 $\pm$   8.58 &  25.81 $\pm$   0.74 &   4.19 $\pm$   0.13 & $ 35.20$ & $ -1.01 \pm   0.07$ &               R\\
DRAO\_J160331.78+530655.15 & 16 03 31.78 $\pm$  0.13 & 53 06 55.15 $\pm$  0.12 &  68.38 $\pm$   1.67 &   3.05 $\pm$   0.51 &   4.45 $\pm$   0.75 & $-61.77$ & $ -1.35 \pm   0.07$ &               R\\
DRAO\_J160333.60+542906.79 & 16 03 33.60 $\pm$  1.00 & 54 29 06.79 $\pm$  3.52 &   8.05 $\pm$   0.41 &   1.07 $\pm$   0.20 &  13.30 $\pm$   2.59 & $ 73.74$ & $ -1.07 \pm   0.21$ &               R\\
DRAO\_J160440.99+543830.91 & 16 04 40.99 $\pm$  0.10 & 54 38 30.91 $\pm$  0.21 &  21.87 $\pm$   0.58 &   2.26 $\pm$   0.18 &  10.34 $\pm$   0.86 & $ 55.57$ & $ -0.91 \pm   0.11$ &               R\\
DRAO\_J160505.45+550045.50 & 16 05 05.45 $\pm$  0.35 & 55 00 45.50 $\pm$  0.52 &  34.21 $\pm$   0.72 &   1.25 $\pm$   0.17 &   3.65 $\pm$   0.50 & $-10.04$ & $ -1.19 \pm   0.08$ &               R\\
DRAO\_J160517.74+532839.61 & 16 05 17.74 $\pm$  0.39 & 53 28 39.61 $\pm$  0.61 &  43.70 $\pm$   0.93 &   2.29 $\pm$   0.22 &   5.24 $\pm$   0.52 & $ 84.33$ & $ -0.78 \pm   0.09$ &               R\\
DRAO\_J160522.34+542927.64 & 16 05 22.34 $\pm$  0.22 & 54 29 27.64 $\pm$  0.48 &   7.72 $\pm$   0.34 &   1.24 $\pm$   0.15 &  16.04 $\pm$   2.02 & $ -9.58$ & $>-0.57$ &               R\\
DRAO\_J160538.09+544146.07 & 16 05 38.09 $\pm$  0.81 & 54 41 46.07 $\pm$  1.13 &   3.06 $\pm$   0.22 &   1.46 $\pm$   0.14 &  47.73 $\pm$   5.82 & $-66.91$ & $>-1.20$ &         \nodata\\
DRAO\_J160538.40+543922.25 & 16 05 38.40 $\pm$  0.09 & 54 39 22.25 $\pm$  0.19 & 218.97 $\pm$   2.86 &  10.24 $\pm$   0.14 &   4.68 $\pm$   0.09 & $-66.91$ & $ -0.89 \pm   0.07$ &               R\\
DRAO\_J160600.00+545406.16 & 16 06 00.00 $\pm$  0.06 & 54 54 06.16 $\pm$  0.13 & 259.46 $\pm$   3.18 &   5.04 $\pm$   0.14 &   1.94 $\pm$   0.06 & $-14.07$ & $ -0.77 \pm   0.07$ &               R\\
DRAO\_J160607.22+534838.16 & 16 06 07.22 $\pm$  1.06 & 53 48 38.16 $\pm$  0.98 &  13.53 $\pm$   0.46 &   1.57 $\pm$   0.16 &  11.62 $\pm$   1.24 & $-10.71$ & $>-0.19$ &               R\\
DRAO\_J160607.30+551608.04 & 16 06 07.30 $\pm$  0.35 & 55 16 08.04 $\pm$  0.78 &   6.30 $\pm$   0.31 &   1.04 $\pm$   0.15 &  16.43 $\pm$   2.46 & $-12.40$ & $ -0.54 \pm   0.53$ &               R\\
DRAO\_J160613.70+550155.81 & 16 06 13.70 $\pm$  1.18 & 55 01 55.81 $\pm$  2.48 &   1.32 $\pm$   0.18 &   0.63 $\pm$   0.12 &  47.68 $\pm$  11.27 & $-16.62$ & $>-1.77$ &         \nodata\\
DRAO\_J160634.97+543455.67 & 16 06 34.97 $\pm$  0.26 & 54 34 55.67 $\pm$  0.53 &  16.92 $\pm$   0.42 &   1.81 $\pm$   0.13 &  10.71 $\pm$   0.79 & $ 51.49$ & $ -0.69 \pm   0.17$ &               R\\
DRAO\_J160721.60+534639.58 & 16 07 21.60 $\pm$  1.35 & 53 46 39.58 $\pm$  1.71 &  15.40 $\pm$   0.49 &   0.56 $\pm$   0.12 &   3.61 $\pm$   0.76 & $-31.90$ & $ -0.41 \pm   0.27$ &               C\\
DRAO\_J160722.80+553059.65 & 16 07 22.80 $\pm$  2.45 & 55 30 59.65 $\pm$  4.03 &  14.26 $\pm$   0.48 &   0.97 $\pm$   0.14 &   6.79 $\pm$   0.98 & $ -8.21$ & $ -0.94 \pm   0.15$ &               R\\
DRAO\_J160724.12+531408.34 & 16 07 24.12 $\pm$  0.19 & 53 14 08.34 $\pm$  0.38 &  10.66 $\pm$   0.49 &   0.92 $\pm$   0.19 &   8.66 $\pm$   1.83 & $-76.36$ & $ -0.49 \pm   0.35$ &               C\\
DRAO\_J160800.77+542150.51 & 16 08 00.77 $\pm$  0.64 & 54 21 50.51 $\pm$  1.89 &   2.52 $\pm$   0.17 &   0.68 $\pm$   0.10 &  27.13 $\pm$   4.39 & $ 13.85$ & $>-1.33$ &         \nodata\\
DRAO\_J160820.78+561355.74 & 16 08 20.78 $\pm$  0.09 & 56 13 55.74 $\pm$  0.22 & 262.79 $\pm$   3.63 &   3.70 $\pm$   0.27 &   1.41 $\pm$   0.11 & $ 33.01$ & $  0.42 \pm   0.08$ &               C\\
DRAO\_J160828.34+541031.91 & 16 08 28.34 $\pm$  1.22 & 54 10 31.91 $\pm$  4.52 &  26.91 $\pm$   0.52 &   0.56 $\pm$   0.10 &   2.09 $\pm$   0.36 & $ 22.36$ & $ -0.47 \pm   0.15$ &               C\\
DRAO\_J160830.91+534421.98 & 16 08 30.91 $\pm$  0.15 & 53 44 21.98 $\pm$  0.33 &  13.27 $\pm$   0.45 &   0.77 $\pm$   0.12 &   5.80 $\pm$   0.95 & $ 71.10$ & $ -0.95 \pm   0.16$ &               R\\
DRAO\_J160838.18+525147.95 & 16 08 38.18 $\pm$  1.55 & 52 51 47.95 $\pm$  2.11 &  26.98 $\pm$   0.98 &   1.64 $\pm$   0.33 &   6.07 $\pm$   1.25 & $ 32.08$ & $ -0.74 \pm   0.12$ &               R\\
DRAO\_J160847.57+561119.82 & 16 08 47.57 $\pm$  0.61 & 56 11 19.82 $\pm$  1.31 &  28.01 $\pm$   0.77 &   1.87 $\pm$   0.22 &   6.66 $\pm$   0.81 & $ 29.40$ & $ -1.16 \pm   0.09$ &               R\\
DRAO\_J160855.82+555640.02 & 16 08 55.82 $\pm$  0.50 & 55 56 40.02 $\pm$  1.91 &   6.64 $\pm$   0.33 &   0.87 $\pm$   0.15 &  13.16 $\pm$   2.35 & $ 64.87$ & $ -1.04 \pm   0.26$ &               R\\
DRAO\_J160903.89+561044.44 & 16 09 03.89 $\pm$  0.93 & 56 10 44.44 $\pm$  1.97 &  13.66 $\pm$   0.52 &   1.87 $\pm$   0.21 &  13.70 $\pm$   1.65 & $ 19.75$ & $ -1.08 \pm   0.14$ &               R\\
DRAO\_J160907.44+535425.60 & 16 09 07.44 $\pm$  0.57 & 53 54 25.60 $\pm$  0.31 &  43.84 $\pm$   0.74 &   0.64 $\pm$   0.11 &   1.47 $\pm$   0.24 & $ 79.98$ & $ -1.01 \pm   0.08$ &               R\\
DRAO\_J160910.22+552632.14 & 16 09 10.22 $\pm$  1.02 & 55 26 32.14 $\pm$  2.00 &   5.76 $\pm$   0.24 &   0.85 $\pm$   0.10 &  14.71 $\pm$   1.88 & $-65.99$ & $ -0.69 \pm   0.47$ &               C\\
DRAO\_J160922.66+561511.20 & 16 09 22.66 $\pm$  0.14 & 56 15 11.20 $\pm$  0.31 &  35.38 $\pm$   0.84 &   2.44 $\pm$   0.24 &   6.91 $\pm$   0.70 & $-33.09$ & $ -0.57 \pm   0.11$ &               R\\
DRAO\_J160929.57+531513.90 & 16 09 29.57 $\pm$  0.75 & 53 15 13.90 $\pm$  0.54 &  33.91 $\pm$   0.77 &   1.51 $\pm$   0.20 &   4.46 $\pm$   0.59 & $-19.09$ & $ -0.26 \pm   0.16$ &               C\\
DRAO\_J160931.01+552502.93 & 16 09 31.01 $\pm$  0.12 & 55 25 02.93 $\pm$  0.30 &  15.83 $\pm$   0.37 &   1.51 $\pm$   0.10 &   9.54 $\pm$   0.68 & $-65.99$ & $ -1.03 \pm   0.12$ &               R\\
DRAO\_J160936.62+552658.88 & 16 09 36.62 $\pm$  0.15 & 55 26 58.88 $\pm$  0.38 &  11.99 $\pm$   0.35 &   1.01 $\pm$   0.10 &   8.43 $\pm$   0.88 & $-65.99$ & $ -0.64 \pm   0.25$ &               R\\
DRAO\_J160944.14+543749.37 & 16 09 44.14 $\pm$  0.91 & 54 37 49.37 $\pm$  2.09 &   9.85 $\pm$   0.30 &   0.39 $\pm$   0.08 &   3.99 $\pm$   0.78 & $ 62.82$ & $>-0.41$ &               R\\
DRAO\_J160953.30+550708.29 & 16 09 53.30 $\pm$  1.69 & 55 07 08.29 $\pm$  2.48 &   0.90 $\pm$   0.11 &   0.37 $\pm$   0.07 &  40.92 $\pm$   9.74 & $ 76.58$ & $>-2.03$ &         \nodata\\
DRAO\_J161002.52+541637.63 & 16 10 02.52 $\pm$  1.23 & 54 16 37.63 $\pm$  3.50 &  16.46 $\pm$   0.36 &   0.47 $\pm$   0.08 &   2.88 $\pm$   0.51 & $-30.87$ & $ -0.28 \pm   0.30$ &               R\\
DRAO\_J161002.88+555243.07 & 16 10 02.88 $\pm$  1.30 & 55 52 43.07 $\pm$  0.26 & 110.05 $\pm$   1.59 &   2.05 $\pm$   0.14 &   1.87 $\pm$   0.13 & $-28.73$ & $ -0.71 \pm   0.07$ &               C\\
DRAO\_J161012.89+544925.90 & 16 10 12.89 $\pm$  0.74 & 54 49 25.90 $\pm$  1.24 &   1.63 $\pm$   0.12 &   0.31 $\pm$   0.07 &  19.11 $\pm$   4.33 & $ 31.66$ & $>-1.63$ &         \nodata\\
DRAO\_J161018.62+531550.44 & 16 10 18.62 $\pm$  0.04 & 53 15 50.44 $\pm$  0.04 & 153.02 $\pm$   2.98 &   1.12 $\pm$   0.19 &   0.73 $\pm$   0.12 & $ 35.22$ & $ -1.05 \pm   0.07$ &               R\\
DRAO\_J161027.53+541246.94 & 16 10 27.53 $\pm$  0.11 & 54 12 46.94 $\pm$  0.25 &  10.36 $\pm$   0.32 &   1.48 $\pm$   0.09 &  14.31 $\pm$   1.01 & $-29.54$ & $ -0.29 \pm   0.46$ &               R\\
DRAO\_J161031.06+540247.83 & 16 10 31.06 $\pm$  1.19 & 54 02 47.83 $\pm$  2.75 &   1.38 $\pm$   0.13 &   0.44 $\pm$   0.09 &  31.99 $\pm$   7.02 & $ 23.08$ & $>-1.74$ &         \nodata\\
DRAO\_J161037.54+532430.78 & 16 10 37.54 $\pm$  0.20 & 53 24 30.78 $\pm$  0.35 &  44.81 $\pm$   0.84 &   0.93 $\pm$   0.16 &   2.08 $\pm$   0.36 & $ 57.79$ & $ -0.63 \pm   0.09$ &               R\\
DRAO\_J161057.53+553528.00 & 16 10 57.53 $\pm$  0.16 & 55 35 28.00 $\pm$  0.36 &  22.87 $\pm$   0.45 &   1.91 $\pm$   0.10 &   8.35 $\pm$   0.46 & $ -3.79$ & $ -0.52 \pm   0.16$ &               C\\
DRAO\_J161100.36+544202.56 & 16 11 00.36 $\pm$  0.31 & 54 42 02.56 $\pm$  0.66 &  37.36 $\pm$   0.60 &   2.03 $\pm$   0.08 &   5.44 $\pm$   0.23 & $-18.84$ & $ -0.58 \pm   0.11$ &               R\\
DRAO\_J161119.85+552846.31 & 16 11 19.85 $\pm$  0.09 & 55 28 46.31 $\pm$  0.20 &  23.71 $\pm$   0.45 &   1.61 $\pm$   0.09 &   6.78 $\pm$   0.40 & $-85.70$ & $ -0.70 \pm   0.13$ &               R\\
DRAO\_J161120.78+543148.25 & 16 11 20.78 $\pm$  0.24 & 54 31 48.25 $\pm$  0.51 &   5.17 $\pm$   0.20 &   0.65 $\pm$   0.08 &  12.47 $\pm$   1.57 & $-77.94$ & $>-0.85$ &               R\\
DRAO\_J161129.78+532709.61 & 16 11 29.78 $\pm$  0.21 & 53 27 09.61 $\pm$  3.53 & 546.91 $\pm$   6.63 &   6.95 $\pm$   0.17 &   1.27 $\pm$   0.03 & $-52.96$ & $ -0.75 \pm   0.07$ &               R\\
DRAO\_J161137.82+555958.24 & 16 11 37.82 $\pm$  0.34 & 55 59 58.24 $\pm$  0.63 &   5.39 $\pm$   0.27 &   0.60 $\pm$   0.12 &  11.16 $\pm$   2.23 & $ 24.47$ & $ -1.07 \pm   0.30$ &         \nodata\\
DRAO\_J161138.18+535924.86 & 16 11 38.18 $\pm$  0.07 & 53 59 24.86 $\pm$  0.11 &  24.04 $\pm$   0.47 &   1.07 $\pm$   0.10 &   4.43 $\pm$   0.43 & $-19.45$ & $ -0.65 \pm   0.14$ &               R\\
DRAO\_J161149.32+550049.82 & 16 11 49.32 $\pm$  1.70 & 55 00 49.82 $\pm$  0.93 &   5.81 $\pm$   0.20 &   0.91 $\pm$   0.08 &  15.59 $\pm$   1.41 & $-76.16$ & $ -0.93 \pm   0.33$ &               R\\
DRAO\_J161159.98+555720.23 & 16 11 59.98 $\pm$  0.25 & 55 57 20.23 $\pm$  0.55 &  15.85 $\pm$   0.42 &   0.51 $\pm$   0.11 &   3.24 $\pm$   0.67 & $ 24.40$ & $ -0.64 \pm   0.19$ &               R\\
DRAO\_J161212.29+552302.18 & 16 12 12.29 $\pm$  0.10 & 55 23 02.18 $\pm$  0.22 & 360.15 $\pm$   4.20 &  21.90 $\pm$   0.07 &   6.08 $\pm$   0.07 & $-29.00$ & $ -1.01 \pm   0.07$ &               R\\
DRAO\_J161223.71+552554.91 & 16 12 23.71 $\pm$  0.46 & 55 25 54.91 $\pm$  0.96 &  12.77 $\pm$   0.30 &   0.75 $\pm$   0.08 &   5.88 $\pm$   0.65 & $-29.00$ & $ -0.37 \pm   0.34$ &               C\\
DRAO\_J161229.02+550636.72 & 16 12 29.02 $\pm$  1.60 & 55 06 36.72 $\pm$  6.23 &  14.36 $\pm$   0.31 &   0.36 $\pm$   0.07 &   2.49 $\pm$   0.46 & $-65.15$ & $ -0.89 \pm   0.16$ &               C\\
DRAO\_J161235.16+562827.98 & 16 12 35.16 $\pm$  0.06 & 56 28 27.98 $\pm$  0.13 & 198.01 $\pm$   4.03 &  12.91 $\pm$   0.26 &   6.52 $\pm$   0.19 & $-40.06$ & $ -0.88 \pm   0.07$ &               R\\
DRAO\_J161249.32+550239.05 & 16 12 49.32 $\pm$  0.83 & 55 02 39.05 $\pm$  1.26 &  10.53 $\pm$   0.26 &   0.36 $\pm$   0.06 &   3.37 $\pm$   0.62 & $ 62.07$ & $ -0.53 \pm   0.33$ &               R\\
DRAO\_J161250.59+560358.07 & 16 12 50.59 $\pm$  0.49 & 56 03 58.07 $\pm$  1.03 &   3.32 $\pm$   0.22 &   0.69 $\pm$   0.12 &  20.77 $\pm$   3.99 & $-26.86$ & $>-1.15$ &               R\\
DRAO\_J161303.24+543224.47 & 16 13 03.24 $\pm$  1.43 & 54 32 24.47 $\pm$  1.87 &  10.34 $\pm$   0.26 &   0.83 $\pm$   0.08 &   8.07 $\pm$   0.76 & $ 64.59$ & $ -0.54 \pm   0.33$ &               R\\
DRAO\_J161316.99+560817.27 & 16 13 16.99 $\pm$  0.38 & 56 08 17.27 $\pm$  0.81 &   5.23 $\pm$   0.28 &   0.78 $\pm$   0.14 &  14.86 $\pm$   2.72 & $ 19.59$ & $>-0.84$ &               R\\
DRAO\_J161320.50+541636.80 & 16 13 20.50 $\pm$  0.60 & 54 16 36.80 $\pm$  1.16 &   6.25 $\pm$   0.22 &   1.42 $\pm$   0.09 &  22.76 $\pm$   1.59 & $  6.78$ & $>-0.72$ &               R\\
DRAO\_J161324.86+553930.64 & 16 13 24.86 $\pm$  0.36 & 55 39 30.64 $\pm$  0.74 &   3.50 $\pm$   0.18 &   0.48 $\pm$   0.08 &  13.70 $\pm$   2.46 & $ 22.19$ & $>-1.11$ &               C\\
DRAO\_J161327.84+551545.72 & 16 13 27.84 $\pm$  0.11 & 55 15 45.72 $\pm$  0.23 &  18.35 $\pm$   0.36 &   0.90 $\pm$   0.07 &   4.92 $\pm$   0.42 & $ 79.37$ & $ -0.55 \pm   0.19$ &               R\\
DRAO\_J161330.89+561755.00 & 16 13 30.89 $\pm$  1.32 & 56 17 55.00 $\pm$  3.85 &  18.30 $\pm$   0.62 &   0.89 $\pm$   0.17 &   4.86 $\pm$   0.93 & $ 78.61$ & $ -0.50 \pm   0.21$ &               R\\
DRAO\_J161331.25+542718.14 & 16 13 31.25 $\pm$  0.14 & 54 27 18.14 $\pm$  0.30 &  99.58 $\pm$   1.35 &   2.34 $\pm$   0.08 &   2.35 $\pm$   0.09 & $-14.46$ & $ -0.70 \pm   0.07$ &               C\\
DRAO\_J161335.74+541103.01 & 16 13 35.74 $\pm$  0.53 & 54 11 03.01 $\pm$  1.14 &   2.76 $\pm$   0.16 &   0.46 $\pm$   0.08 &  16.82 $\pm$   3.12 & $-68.32$ & $>-1.27$ &         \nodata\\
DRAO\_J161341.04+524911.86 & 16 13 41.04 $\pm$  0.23 & 52 49 11.86 $\pm$  1.11 &  53.21 $\pm$   1.79 &   7.79 $\pm$   0.57 &  14.64 $\pm$   1.17 & $-80.12$ & $ -0.71 \pm   0.09$ &               R\\
DRAO\_J161341.76+561215.37 & 16 13 41.76 $\pm$  0.53 & 56 12 15.37 $\pm$  1.67 &  12.96 $\pm$   0.44 &   1.91 $\pm$   0.16 &  14.73 $\pm$   1.34 & $-19.26$ & $ -1.31 \pm   0.12$ &         \nodata\\
DRAO\_J161345.89+561339.40 & 16 13 45.89 $\pm$  1.37 & 56 13 39.40 $\pm$  3.36 &   4.17 $\pm$   0.27 &   1.61 $\pm$   0.17 &  38.61 $\pm$   4.70 & $-19.26$ & $>-0.99$ &               R\\
DRAO\_J161346.99+541410.72 & 16 13 46.99 $\pm$  0.00 & 54 14 10.72 $\pm$  0.00 &  14.26 $\pm$   0.33 &   0.75 $\pm$   0.09 &   5.23 $\pm$   0.61 & $-68.32$ & $ -0.72 \pm   0.19$ &               R\\
DRAO\_J161355.42+545753.14 & 16 13 55.42 $\pm$  1.22 & 54 57 53.14 $\pm$  2.56 &   0.49 $\pm$   0.08 &   0.29 $\pm$   0.06 &  60.25 $\pm$  15.69 & $ 78.72$ & $>-2.45$ &         \nodata\\
DRAO\_J161358.06+550218.49 & 16 13 58.06 $\pm$  1.00 & 55 02 18.49 $\pm$  1.45 &   1.58 $\pm$   0.11 &   0.27 $\pm$   0.06 &  16.75 $\pm$   3.78 & $ 68.22$ & $>-1.65$ &         \nodata\\
DRAO\_J161400.34+535715.08 & 16 14 00.34 $\pm$  0.11 & 53 57 15.08 $\pm$  0.24 &  16.74 $\pm$   0.39 &   1.20 $\pm$   0.11 &   7.16 $\pm$   0.66 & $-33.04$ & $ -0.32 \pm   0.28$ &               R\\
DRAO\_J161406.43+531918.52 & 16 14 06.43 $\pm$  0.00 & 53 19 18.52 $\pm$  0.00 &  22.41 $\pm$   0.75 &   1.47 $\pm$   0.21 &   6.56 $\pm$   0.98 & $ 35.92$ & $ -0.70 \pm   0.14$ &               R\\
DRAO\_J161412.14+554132.75 & 16 14 12.14 $\pm$  0.51 & 55 41 32.75 $\pm$  1.23 &   5.40 $\pm$   0.22 &   0.50 $\pm$   0.08 &   9.26 $\pm$   1.59 & $-67.73$ & $>-0.82$ &               R\\
DRAO\_J161416.70+554319.27 & 16 14 16.70 $\pm$  2.22 & 55 43 19.27 $\pm$  4.69 &   0.69 $\pm$   0.12 &   0.44 $\pm$   0.08 &  63.52 $\pm$  15.95 & $-55.34$ & $>-2.21$ &         \nodata\\
DRAO\_J161421.62+553651.26 & 16 14 21.62 $\pm$  0.38 & 55 36 51.26 $\pm$  1.13 &  42.37 $\pm$   0.67 &   3.19 $\pm$   0.09 &   7.53 $\pm$   0.24 & $-71.73$ & $ -0.62 \pm   0.10$ &               R\\
DRAO\_J161435.47+553841.57 & 16 14 35.47 $\pm$  3.01 & 55 38 41.57 $\pm$  5.14 &   0.73 $\pm$   0.11 &   0.56 $\pm$   0.08 &  76.18 $\pm$  16.18 & $-71.73$ & $>-2.17$ &         \nodata\\
DRAO\_J161509.94+531015.82 & 16 15 09.94 $\pm$  0.13 & 53 10 15.82 $\pm$  0.27 & 100.57 $\pm$   1.79 &   4.44 $\pm$   0.36 &   4.42 $\pm$   0.36 & $-15.02$ & $ -0.69 \pm   0.07$ &               R\\
DRAO\_J161527.67+542712.46 & 16 15 27.67 $\pm$  0.67 & 54 27 12.46 $\pm$  2.04 &   7.57 $\pm$   0.26 &   1.01 $\pm$   0.08 &  13.38 $\pm$   1.16 & $ -4.49$ & $>-0.59$ &               R\\
DRAO\_J161530.60+545234.86 & 16 15 30.60 $\pm$  0.65 & 54 52 34.86 $\pm$  1.21 &   6.98 $\pm$   0.24 &   0.70 $\pm$   0.07 &  10.01 $\pm$   1.06 & $ 67.00$ & $ -0.99 \pm   0.26$ &         \nodata\\
DRAO\_J161537.82+534646.96 & 16 15 37.82 $\pm$  0.21 & 53 46 46.96 $\pm$  0.58 &  60.76 $\pm$   0.99 &   2.51 $\pm$   0.14 &   4.14 $\pm$   0.24 & $-22.18$ & $ -0.77 \pm   0.08$ &               R\\
DRAO\_J161547.14+532820.21 & 16 15 47.14 $\pm$  1.18 & 53 28 20.21 $\pm$  3.00 &   3.69 $\pm$   0.30 &   1.03 $\pm$   0.19 &  28.02 $\pm$   5.71 & $-72.79$ & $>-1.08$ &               C\\
DRAO\_J161549.70+551647.78 & 16 15 49.70 $\pm$  0.71 & 55 16 47.78 $\pm$  0.23 &  30.10 $\pm$   0.50 &   1.75 $\pm$   0.07 &   5.80 $\pm$   0.26 & $-51.74$ & $ -0.63 \pm   0.12$ &               R\\
DRAO\_J161558.82+552505.48 & 16 15 58.82 $\pm$  1.12 & 55 25 05.48 $\pm$  5.10 &   0.68 $\pm$   0.10 &   0.35 $\pm$   0.07 &  51.34 $\pm$  12.52 & $ 61.25$ & $>-2.22$ &         \nodata\\
DRAO\_J161559.21+532416.56 & 16 15 59.21 $\pm$  0.16 & 53 24 16.56 $\pm$  0.33 & 217.89 $\pm$   3.04 &  13.26 $\pm$   0.24 &   6.09 $\pm$   0.14 & $-73.03$ & $ -0.74 \pm   0.07$ &               R\\
DRAO\_J161603.82+540434.46 & 16 16 03.82 $\pm$  0.35 & 54 04 34.46 $\pm$  0.71 &   4.65 $\pm$   0.22 &   0.48 $\pm$   0.09 &  10.22 $\pm$   2.04 & $-17.16$ & $>-0.92$ &               C\\
DRAO\_J161616.49+541724.65 & 16 16 16.49 $\pm$  0.22 & 54 17 24.65 $\pm$  0.48 &   5.96 $\pm$   0.23 &   0.50 $\pm$   0.08 &   8.33 $\pm$   1.43 & $-79.66$ & $ -0.85 \pm   0.36$ &               R\\
DRAO\_J161623.35+545739.71 & 16 16 23.35 $\pm$  0.09 & 54 57 39.71 $\pm$  0.19 &  12.39 $\pm$   0.28 &   0.48 $\pm$   0.07 &   3.86 $\pm$   0.56 & $-75.24$ & $ -0.84 \pm   0.18$ &               C\\
DRAO\_J161623.93+552703.82 & 16 16 23.93 $\pm$  1.79 & 55 27 03.82 $\pm$  0.91 &  12.10 $\pm$   0.29 &   0.97 $\pm$   0.08 &   8.03 $\pm$   0.69 & $ 61.25$ & $ -0.95 \pm   0.17$ &               R\\
DRAO\_J161639.14+562032.64 & 16 16 39.14 $\pm$  1.19 & 56 20 32.64 $\pm$  2.05 &  18.71 $\pm$   0.63 &   1.33 $\pm$   0.20 &   7.10 $\pm$   1.07 & $ 52.69$ & $ -0.52 \pm   0.20$ &               R\\
DRAO\_J161639.46+554525.74 & 16 16 39.46 $\pm$  0.11 & 55 45 25.74 $\pm$  0.20 &  72.72 $\pm$   1.63 &   1.00 $\pm$   0.10 &   1.38 $\pm$   0.14 & $-38.96$ & $ -0.63 \pm   0.08$ &               C\\
DRAO\_J161640.42+535813.26 & 16 16 40.42 $\pm$  0.32 & 53 58 13.26 $\pm$  0.54 & 355.62 $\pm$   4.41 &   3.94 $\pm$   0.12 &   1.11 $\pm$   0.04 & $ 53.70$ & $ -0.84 \pm   0.07$ &               R\\
DRAO\_J161647.42+535951.00 & 16 16 47.42 $\pm$  1.08 & 53 59 51.00 $\pm$  3.82 &   3.35 $\pm$   0.20 &   0.48 $\pm$   0.10 &  14.30 $\pm$   3.08 & $ 53.70$ & $ -1.83 \pm   0.19$ &               R\\
DRAO\_J161757.12+545111.41 & 16 17 57.12 $\pm$  1.48 & 54 51 11.41 $\pm$ 10.96 &  15.57 $\pm$   0.33 &   0.94 $\pm$   0.08 &   6.04 $\pm$   0.52 & $-74.29$ & $ -1.07 \pm   0.12$ &               R\\
DRAO\_J161758.06+543518.64 & 16 17 58.06 $\pm$  0.19 & 54 35 18.64 $\pm$  0.38 &   7.52 $\pm$   0.26 &   0.33 $\pm$   0.07 &   4.41 $\pm$   0.93 & $-14.79$ & $>-0.59$ &               C\\
DRAO\_J161801.18+542657.26 & 16 18 01.18 $\pm$  0.00 & 54 26 57.26 $\pm$  0.00 &  12.35 $\pm$   0.31 &   0.43 $\pm$   0.08 &   3.45 $\pm$   0.66 & $-63.62$ & $ -0.53 \pm   0.28$ &               C\\
DRAO\_J161807.58+544237.87 & 16 18 07.58 $\pm$  1.41 & 54 42 37.87 $\pm$  3.44 &   8.11 $\pm$   0.27 &   0.68 $\pm$   0.08 &   8.42 $\pm$   1.02 & $-19.82$ & $>-0.54$ &               R\\
DRAO\_J161826.28+542532.77 & 16 18 26.28 $\pm$  0.00 & 54 25 32.77 $\pm$  0.00 &  13.55 $\pm$   0.33 &   0.51 $\pm$   0.09 &   3.76 $\pm$   0.65 & $-67.46$ & $ -0.44 \pm   0.29$ &               R\\
DRAO\_J161831.99+543834.58 & 16 18 31.99 $\pm$  0.00 & 54 38 34.58 $\pm$  0.00 &   7.69 $\pm$   0.26 &   0.52 $\pm$   0.08 &   6.71 $\pm$   1.07 & $-23.76$ & $>-0.58$ &               C\\
DRAO\_J161833.00+543145.77 & 16 18 33.00 $\pm$  0.69 & 54 31 45.77 $\pm$  0.46 &  58.09 $\pm$   1.81 &   4.67 $\pm$   0.09 &   8.04 $\pm$   0.29 & $  5.05$ & $ -0.66 \pm   0.09$ &               R\\
DRAO\_J161859.40+545239.68 & 16 18 59.40 $\pm$  0.57 & 54 52 39.68 $\pm$  1.84 &  38.61 $\pm$   0.63 &   1.16 $\pm$   0.08 &   3.00 $\pm$   0.23 & $-72.51$ & $ -0.53 \pm   0.11$ &               R\\
DRAO\_J161900.70+542938.18 & 16 19 00.70 $\pm$  0.01 & 54 29 38.18 $\pm$  0.05 & 233.81 $\pm$   3.11 &   2.69 $\pm$   0.10 &   1.15 $\pm$   0.04 & $ 34.91$ & $ -1.01 \pm   0.07$ &               R\\
DRAO\_J161915.31+550513.27 & 16 19 15.31 $\pm$  3.59 & 55 05 13.27 $\pm$  9.10 &   5.60 $\pm$   0.21 &   0.82 $\pm$   0.08 &  14.62 $\pm$   1.61 & $ 82.66$ & $ -1.04 \pm   0.29$ &               R\\
DRAO\_J161919.97+553600.97 & 16 19 19.97 $\pm$  0.39 & 55 36 00.97 $\pm$  3.71 &  55.89 $\pm$   1.17 &   2.10 $\pm$   0.10 &   3.76 $\pm$   0.20 & $-70.93$ & $ -0.41 \pm   0.10$ &               R\\
DRAO\_J161920.23+544816.99 & 16 19 20.23 $\pm$  0.19 & 54 48 16.99 $\pm$  0.37 &  67.59 $\pm$   1.34 &   3.23 $\pm$   0.09 &   4.77 $\pm$   0.16 & $-78.65$ & $ -0.56 \pm   0.08$ &               C\\
DRAO\_J161923.38+540032.98 & 16 19 23.38 $\pm$  1.38 & 54 00 32.98 $\pm$  3.93 &  41.01 $\pm$   0.76 &   2.94 $\pm$   0.15 &   7.17 $\pm$   0.40 & $-54.27$ & $ -0.73 \pm   0.09$ &               R\\
DRAO\_J161924.79+555113.57 & 16 19 24.79 $\pm$  0.51 & 55 51 13.57 $\pm$  1.13 &   2.92 $\pm$   0.20 &   0.97 $\pm$   0.12 &  33.18 $\pm$   4.77 & $-81.23$ & $>-1.23$ &               R\\
DRAO\_J162006.53+543233.11 & 16 20 06.53 $\pm$  0.19 & 54 32 33.11 $\pm$  0.41 &   6.58 $\pm$   0.25 &   0.63 $\pm$   0.10 &   9.64 $\pm$   1.53 & $ 21.66$ & $>-0.68$ &               R\\
DRAO\_J162008.95+560008.68 & 16 20 08.95 $\pm$  0.14 & 56 00 08.68 $\pm$  0.32 &  35.14 $\pm$   0.72 &   0.90 $\pm$   0.15 &   2.56 $\pm$   0.43 & $-81.54$ & $  0.23 \pm   0.30$ &               C\\
DRAO\_J162011.28+562615.43 & 16 20 11.28 $\pm$  0.41 & 56 26 15.43 $\pm$  0.95 &  20.55 $\pm$   0.78 &   1.45 $\pm$   0.29 &   7.05 $\pm$   1.43 & $ 13.34$ & $ -0.52 \pm   0.19$ &               C\\
DRAO\_J162038.02+545129.41 & 16 20 38.02 $\pm$  0.20 & 54 51 29.41 $\pm$  0.42 &  11.00 $\pm$   0.34 &   0.75 $\pm$   0.10 &   6.83 $\pm$   0.89 & $ 57.42$ & $ -0.92 \pm   0.19$ &               R\\
DRAO\_J162040.51+551743.33 & 16 20 40.51 $\pm$  0.42 & 55 17 43.33 $\pm$  0.92 &   2.75 $\pm$   0.17 &   0.40 $\pm$   0.08 &  14.46 $\pm$   3.19 & $-74.46$ & $>-1.27$ &               R\\
DRAO\_J162047.54+545037.25 & 16 20 47.54 $\pm$  0.76 & 54 50 37.25 $\pm$  1.34 &   7.91 $\pm$   0.28 &   0.43 $\pm$   0.09 &   5.41 $\pm$   1.11 & $ 57.42$ & $ -0.97 \pm   0.24$ &               R\\
DRAO\_J162055.06+544854.79 & 16 20 55.06 $\pm$  0.22 & 54 48 54.79 $\pm$  0.48 &  65.52 $\pm$   1.98 &   0.75 $\pm$   0.10 &   1.14 $\pm$   0.16 & $ 57.42$ & $  0.80 \pm   0.37$ &               C\\
DRAO\_J162114.23+552349.42 & 16 21 14.23 $\pm$  0.15 & 55 23 49.42 $\pm$  0.33 &  63.01 $\pm$   1.39 &   2.68 $\pm$   0.12 &   4.25 $\pm$   0.21 & $-48.98$ & $ -0.89 \pm   0.08$ &               R\\
DRAO\_J162116.20+551758.27 & 16 21 16.20 $\pm$  1.88 & 55 17 58.27 $\pm$  0.80 &  15.35 $\pm$   0.38 &   0.86 $\pm$   0.11 &   5.59 $\pm$   0.71 & $-58.81$ & $ -0.61 \pm   0.21$ &               R\\
DRAO\_J162119.30+553834.22 & 16 21 19.30 $\pm$  0.29 & 55 38 34.22 $\pm$  1.00 &  11.63 $\pm$   0.40 &   0.92 $\pm$   0.13 &   7.95 $\pm$   1.13 & $ 81.47$ & $ -0.73 \pm   0.23$ &               R\\
DRAO\_J162142.84+540353.82 & 16 21 42.84 $\pm$  0.04 & 54 03 53.82 $\pm$  0.09 &  47.40 $\pm$   0.92 &   2.00 $\pm$   0.20 &   4.21 $\pm$   0.42 & $-83.46$ & $ -0.46 \pm   0.10$ &               C\\
DRAO\_J162145.50+542111.34 & 16 21 45.50 $\pm$  1.22 & 54 21 11.34 $\pm$  2.59 &   2.24 $\pm$   0.20 &   0.66 $\pm$   0.13 &  29.42 $\pm$   6.29 & $ 29.28$ & $>-1.41$ &               R\\
DRAO\_J162145.98+554950.20 & 16 21 45.98 $\pm$  0.87 & 55 49 50.20 $\pm$  2.14 &  34.85 $\pm$   0.72 &   2.07 $\pm$   0.17 &   5.94 $\pm$   0.50 & $ 53.71$ & $ -0.55 \pm   0.12$ &               R\\
DRAO\_J162209.53+552341.46 & 16 22 09.53 $\pm$  0.00 & 55 23 41.46 $\pm$  0.00 &  39.63 $\pm$   0.71 &   1.63 $\pm$   0.13 &   4.12 $\pm$   0.34 & $-12.31$ & $ -0.87 \pm   0.09$ &               R\\
DRAO\_J162209.89+552527.77 & 16 22 09.89 $\pm$  0.65 & 55 25 27.77 $\pm$  1.45 &  39.46 $\pm$   0.71 &   2.82 $\pm$   0.13 &   7.15 $\pm$   0.36 & $-12.31$ & $ -0.61 \pm   0.10$ &               R\\
DRAO\_J162216.46+552209.59 & 16 22 16.46 $\pm$  0.57 & 55 22 09.59 $\pm$  1.32 &  20.02 $\pm$   0.47 &   0.86 $\pm$   0.12 &   4.29 $\pm$   0.63 & $-12.31$ & $ -0.79 \pm   0.14$ &               C\\
DRAO\_J162224.67+543838.08 & 16 22 24.67 $\pm$  0.87 & 54 38 38.08 $\pm$  2.51 &  13.85 $\pm$   0.46 &   0.63 $\pm$   0.12 &   4.53 $\pm$   0.88 & $ 12.80$ & $ -0.25 \pm   0.37$ &               R\\
DRAO\_J162233.58+553035.14 & 16 22 33.58 $\pm$  0.09 & 55 30 35.14 $\pm$  0.20 &  24.68 $\pm$   0.56 &   0.73 $\pm$   0.13 &   2.96 $\pm$   0.55 & $  9.92$ & $ -0.56 \pm   0.15$ &               R\\
DRAO\_J162234.03+551300.41 & 16 22 34.03 $\pm$  0.41 & 55 13 00.41 $\pm$  0.86 &  49.97 $\pm$   0.85 &   1.26 $\pm$   0.13 &   2.53 $\pm$   0.26 & $ 27.84$ & $ -0.48 \pm   0.10$ &               R\\
DRAO\_J162321.72+541418.78 & 16 23 21.72 $\pm$  1.88 & 54 14 18.78 $\pm$  3.83 &  14.81 $\pm$   0.54 &   1.68 $\pm$   0.21 &  11.35 $\pm$   1.50 & $-22.78$ & $ -0.48 \pm   0.26$ &               R\\
DRAO\_J162332.64+553127.66 & 16 23 32.64 $\pm$  1.04 & 55 31 27.66 $\pm$  2.44 &   7.81 $\pm$   0.37 &   0.76 $\pm$   0.15 &   9.78 $\pm$   2.01 & $-59.91$ & $ -1.04 \pm   0.22$ &               R\\
DRAO\_J162346.85+553208.63 & 16 23 46.85 $\pm$  1.38 & 55 32 08.63 $\pm$  2.01 &   8.04 $\pm$   0.38 &   0.77 $\pm$   0.16 &   9.57 $\pm$   2.03 & $-26.76$ & $ -0.51 \pm   0.44$ &               C\\
DRAO\_J162500.26+555020.58 & 16 25 00.26 $\pm$  0.43 & 55 50 20.58 $\pm$  0.92 &  27.70 $\pm$   0.94 &   1.41 $\pm$   0.29 &   5.10 $\pm$   1.06 & $-14.96$ & $ -0.77 \pm   0.12$ &               R\\
DRAO\_J162513.85+555634.33 & 16 25 13.85 $\pm$  0.52 & 55 56 34.33 $\pm$  0.79 &  11.76 $\pm$   0.67 &   3.24 $\pm$   0.42 &  27.56 $\pm$   3.91 & $-11.43$ & $ -0.78 \pm   0.23$ &         \nodata\\
DRAO\_J162513.99+560032.00 & 16 25 13.99 $\pm$  1.94 & 56 00 32.00 $\pm$  3.41 &   2.50 $\pm$   0.40 &   1.82 $\pm$   0.38 &  72.69 $\pm$  19.27 & $-13.23$ & $>-1.34$ &               R\\
DRAO\_J162525.97+555658.67 & 16 25 25.97 $\pm$  1.30 & 55 56 58.67 $\pm$  2.26 &  11.77 $\pm$   0.69 &   1.82 $\pm$   0.38 &  15.42 $\pm$   3.35 & $-11.43$ & $ -1.07 \pm   0.17$ &               R\\
DRAO\_J162544.02+550833.97 & 16 25 44.02 $\pm$  0.33 & 55 08 33.97 $\pm$  0.28 &  19.97 $\pm$   0.69 &   1.99 $\pm$   0.25 &   9.96 $\pm$   1.31 & $-72.67$ & $ -0.92 \pm   0.12$ &               R\\
DRAO\_J162602.54+553858.16 & 16 26 02.54 $\pm$  0.93 & 55 38 58.16 $\pm$  1.80 &  17.94 $\pm$   0.77 &   1.48 $\pm$   0.31 &   8.25 $\pm$   1.79 & $-16.31$ & $ -0.77 \pm   0.16$ &               R\\
DRAO\_J162607.27+550300.76 & 16 26 07.27 $\pm$  0.13 & 55 03 00.76 $\pm$  0.29 &  45.55 $\pm$   1.03 &   1.86 $\pm$   0.28 &   4.08 $\pm$   0.62 & $-10.49$ & $ -0.70 \pm   0.09$ &               C\\
DRAO\_J162611.11+545219.16 & 16 26 11.11 $\pm$  1.98 & 54 52 19.16 $\pm$  3.88 &  30.77 $\pm$   0.98 &   2.34 $\pm$   0.30 &   7.61 $\pm$   1.01 & $ 19.56$ & $ -0.81 \pm   0.11$ &               R\\
DRAO\_J162634.13+544206.88 & 16 26 34.13 $\pm$  0.51 & 54 42 06.88 $\pm$  1.01 &  85.47 $\pm$   1.63 &   6.79 $\pm$   0.36 &   7.94 $\pm$   0.45 & $-19.51$ & $ -0.21 \pm   0.09$ &               C\\
\enddata
\tablenotetext{*}{C = object classified as FIRST compact source, R = object classified as FIRST resolved source.}
\label{catalog}
\end{deluxetable}


\begin{deluxetable}{rrr}
\tablecaption{DRAO ELAIS N1 Euclidean-normalized polarized source counts.}
\tablehead{
\colhead{$P_{\rm 0}$} & \colhead{$N$} & \colhead{$P_{\rm 0}^{2.5}dN/dP$}\\
\colhead{(mJy)} & \colhead{ } & \colhead{(Jy$^{1.5}$sr$^{-1}$)}}
\startdata
0.45 & 17 & 0.84 $\pm$ 0.13\\
0.67 & 23 & 0.81 $\pm$ 0.07\\
0.94 & 25 & 0.98 $\pm$ 0.08\\
1.43 & 20 & 1.15 $\pm$ 0.10\\
1.98 & 20 & 1.52 $\pm$ 0.11\\
2.86 & 10 & 1.07 $\pm$ 0.12\\
4.37 & 5 & 0.95 $\pm$ 0.14\\
6.87 & 2 & 0.80 $\pm$ 0.18\\
8.97 & 2 & 1.03 $\pm$ 0.25\\
13.09 & 2 & 1.76 $\pm$ 0.38\\
21.90 & 1 & 2.20 $\pm$ 0.60\\
25.81 & 1 & 2.28 $\pm$ 0.53\\
\enddata
\label{polcounts}
\end{deluxetable}


\begin{deluxetable}{rrrr}
\tablecaption{Number of DRAO ELAIS N1 sources with a match in the WENSS catalogue with $S_{1.4} \ge 1.19\,$mJy.}
\tablehead{
\colhead{ } & \colhead{All Sources} & \colhead{Polarized} & \colhead{No Polarization}
}
\startdata
DRAO sources & 751 & 131 & 620\\
WENSS matches & 343 & 103 & 240\\
Percentage & $46\,\pm\,3\%$ & $79\,\pm\,8\%$ & $39\,\pm\,3\%$\\
\enddata
\label{wensstable}
\end{deluxetable}

\begin{deluxetable}{lrrrr}
\tablecaption{Classification of DRAO sources with FIRST counterparts.}
\tablehead{
\colhead{} & \colhead{Compact} & \colhead{Percentage} & \colhead{Resolved} & \colhead{Percentage}
}
\startdata
No detectable polarization & 347 & 62 $\pm$ 3\% & 215 & 38 $\pm$ 3\%\\
Polarized Sources & 27 & 23 $\pm$ 5\% & 88 & 77 $\pm$ 8\%\\
All sources & 374 & 55 $\pm$ 3\% & 303 & 45 $\pm$ 3\%\\
\enddata
\label{firstpercent}
\end{deluxetable}




\end{document}